\documentclass[aip, pof,graphicx]{revtex4-2}

\usepackage{graphicx}
\usepackage{dcolumn}
\usepackage{bm}
\usepackage[utf8]{inputenc}
\usepackage[T1]{fontenc}
\usepackage{mathptmx}
\usepackage{color}
\usepackage{longtable}
\usepackage{changes}
\usepackage{epstopdf} 

\draft 

\begin{document}


\title[Motion and clustering of bonded particles in narrow solid-liquid fluidized beds]{Motion and clustering of bonded particles in narrow solid-liquid fluidized beds\\
This article may be downloaded for personal use only. Any other use requires prior permission of the author and AIP Publishing. This article appeared in Phys. Fluids 33, 023303 (2021) and may be found at https://doi.org/10.1063/5.0035718.} 



\author{Fernando David C\'u\~nez}
\author{Nicolao Cerqueira Lima}
\author{Erick M. Franklin}%
 \email{erick.franklin@unicamp.br}
 \thanks{Corresponding author}
\affiliation{ 
School of Mechanical Engineering, UNICAMP - University of Campinas,\\
Rua Mendeleyev, 200, Campinas, SP, Brazil
}%


\date{\today}

\begin{abstract}
This paper presents an experimental and numerical investigation of solid-liquid fluidized beds consisting of bonded spheres in very narrow tubes, i.e., when the ratio between the tube and grain diameters is small. In narrow beds, high confinement effects have proved to induce crystallization, jamming and different patterns, which can be intensified or modified if some grains are bonded together. In order to investigate that, we produced duos and trios of bonded aluminum spheres with diameter of 4.8 mm, and formed beds consisting either of 150-300 duos or 100-200 trios in a 25.4 mm-ID pipe, which were submitted to water velocities above those necessary for fluidization. For the experiments, we filmed the bed with high-speed and conventional cameras and processed the images, obtaining measurements at both the bed and grain scales. For the numerical part, we computed the bed evolution for the same conditions with a CFD-DEM (computational fluid dynamics - discrete element method) code. Our results show distinct motions for individual duos and trios, and different structures within the bed. We also found that jamming may occur suddenly for trios, where even the microscopic motion (fluctuation at the grain scale) stops, calling into question the fluidization conditions for those cases.
\end{abstract}


\maketitle 

\section{INTRODUCTION}

A common way to enhance heat and mass transfers between fluid and solids is by increasing their contact area and relative velocities, which is typically achieved by fragmenting the solid into grains and imposing an ascending fluid flow through them. In practical applications, in order to maximize transfers, flow velocities are adjusted to maintain a granular bed suspended in a tube section, forming a fluidized bed. Fluidized beds are thus frequently found in industry for their high transfer rates and simple construction. However, although of straightforward architecture, fluidized beds have a rich dynamics in which different scales are involved. While fluid forces and solid contacts are important at the grain scale, contact networks, clusters, plugs, crystals and other structures are found at bed (or tube) scale.

In addition to having diverse scales in a given bed, the bed dynamics differs according to the fluid state (gas or liquid) and the problem is not scale invariant. For gas-solid fluidized beds, particle collisions and fluid drag are usually recognized as the most pertinent mechanisms \cite{koch,Sundaresan}, while for solid-liquid fluidized beds (SLFBs), virtual mass and pressure forces can be as important as fluid drag and particle collisions \cite{Cunez,Cunez2}. For large beds, instabilities in the form of bubbles and slugs usually appear \cite{Geldart,Sundaresan,guazzelli_book,Koralkar}, while for narrow beds, those with thickness between 10 to 100 grain diameters, transverse waves, blobs and bubbles appear \cite{Duru,Duru2}, and for very narrow beds, those whose thicknesses are smaller than 10 grain diameters, plugs, crystallization and jamming may occur \cite{Cunez,Cunez2,Cunez3}.

Narrow SLFBs have been studied over the last decades for their different dynamics when compared to large beds \cite{Anderson,ElKaissy,Didwania,Zenit,Zenit2,Duru2,Duru,Aguilar,Ghatage,Cunez,Cunez2,Cunez3}. In particular, C\'u\~nez and Franklin \cite{Cunez,Cunez2} investigated the behavior of very narrow beds for which ratios between the tube diameter $D$ and that of grains $d$ were smaller than 6, conducting both experiments and CFD-DEM (computational fluid dynamics - discrete element method) computations. They showed that alternating regions of high and low particle fractions in the form of granular plugs appear due to a dense network of contact forces that goes from the bed center line to the tube wall, evidencing the effects of high confinement on the bed dynamics. More recently, C\'u\~nez and Franklin \cite{Cunez3} investigated experimentally very narrow SLFBs consisting of alumina and glass beads and in which $D/d$ = 4.2 and 3.2, for decreasing water flows that reached velocities still higher than that for incipient fluidization and were afterward increased by small steps. They showed that crystallization and jamming appear in some cases, and that different structures appear depending on the grain type, calling into question the fluidization conditions when confinement is strong.

Narrow and very narrow beds are encountered in biological processes involving mass transfers, where in some cases an organic film grows around each solid particle \cite{Dempsey}. For example, in the biological treatment of domestic wastewater, fluidized-bed bioreactors have been employed over the last decades in regions where extensive treatment of wastewater is not accessible \cite{Dempsey,Nelson}. Those reactors are usually small and, depending on the size of sediments, may be classified as narrow, something that is intensified by the bonding of particles due to the adhesion properties of the organic film. The bonding of two or more particles reduces the exchange surface, decreasing the bioreactor efficiency, and may even cause blockages within the bed and prevent the reactor operation.

Besides the practical applications of fluidized beds with bonded particles, understanding them is important to gain physical insight on the behavior of beds containing irregular shaped grains. Recent studies have shown that the geometry and properties of particles can induce particle agglomeration and have an effect on grain segregation in polydispersed beds \cite{Shiyuan,Zhou2,Boer}, and that spouted and fluidized beds consisting of cohesive spheres, which may bond together due to adhesion forces, behave differently from those of loose spheres \cite{Mikami,Nascimento,Xu,Zhou3,Wang2}. In general, given the irregular shape of bonded structures, the bonded elements tend to form clusters more easily that, when large enough, interact with walls \cite{Mikami}. 

This paper presents an experimental and numerical investigation of SLFBs consisting of bonded grains in very narrow tubes. To form each bed element, either two or three aluminum spheres with diameter of 4.8 mm were bonded together, forming duos and trios, respectively, and the fluidized beds consisted either of 150-300 duos or 100-200 trios in a 25.4 mm-ID pipe, submitted to water velocities above those necessary for fluidization. The experiments were performed in a transparent tube, where we filmed the bed with high-speed and conventional cameras and obtained measurements at both the bed and grain scales by using image processing. The numerical computations were conducted using an open-source CFD-DEM code, which computed the bed evolution for the same conditions used in the experiments. Our results show distinct motions for individual duos and trios, and the existence of different structures within the bed. We also found that jamming may occur suddenly for trios, where even the microscopic motion (fluctuation at the grain scale) stops. The present results call into question the fluidization conditions for those cases and provide new insights for clustering and jamming in SLFBs.

The next sections present the experimental and numerical setups, and are followed by a section containing the results and discussions. The last section presents the conclusions.

\section{EXPERIMENTAL SETUP}

\begin{figure}[ht]
	\centering
	\includegraphics[width=0.95\columnwidth]{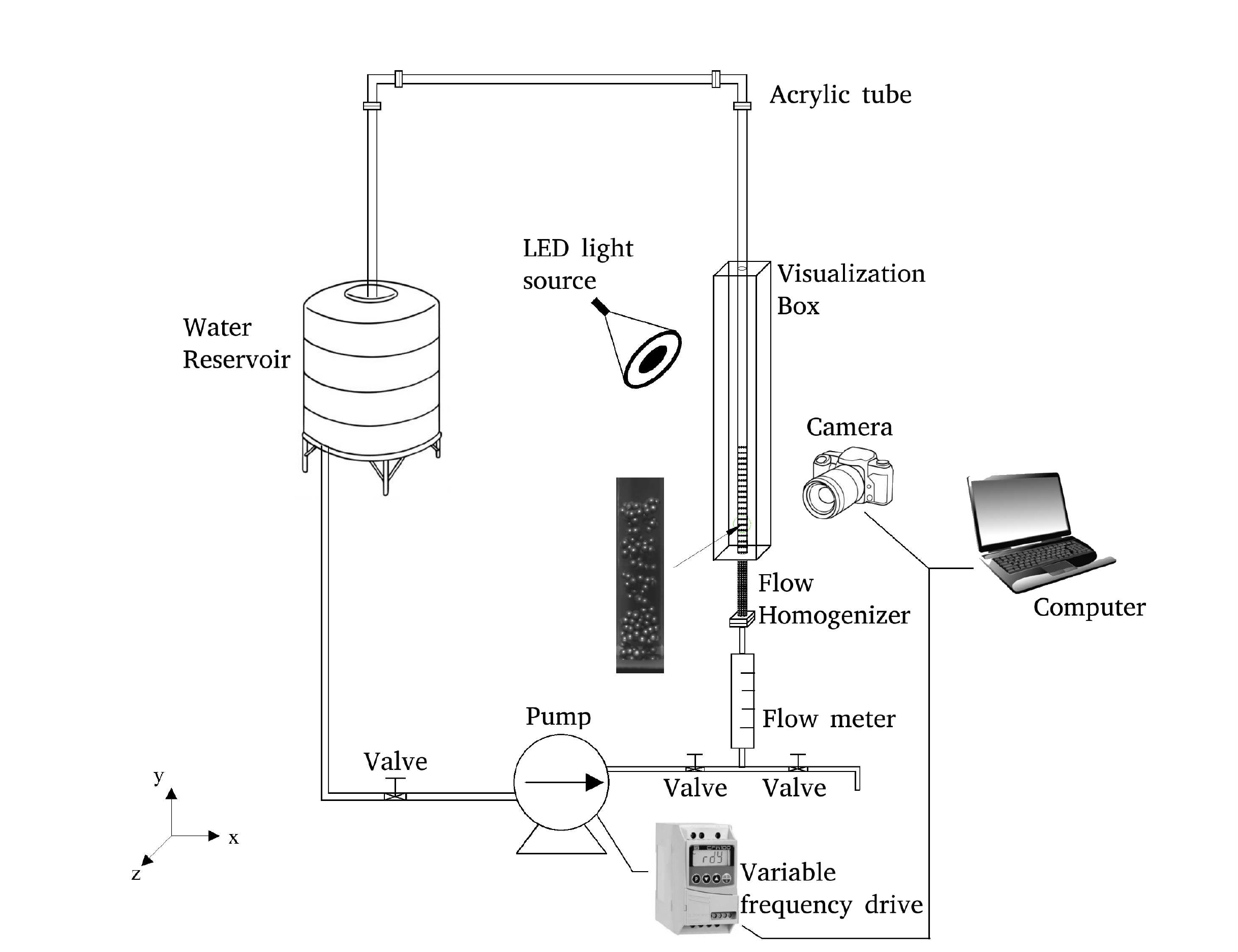}
	\caption{Layout of the experimental setup.}
	\label{fig:1}
\end{figure}

\begin{figure}[ht]
	\centering
	\includegraphics[width=0.3\columnwidth]{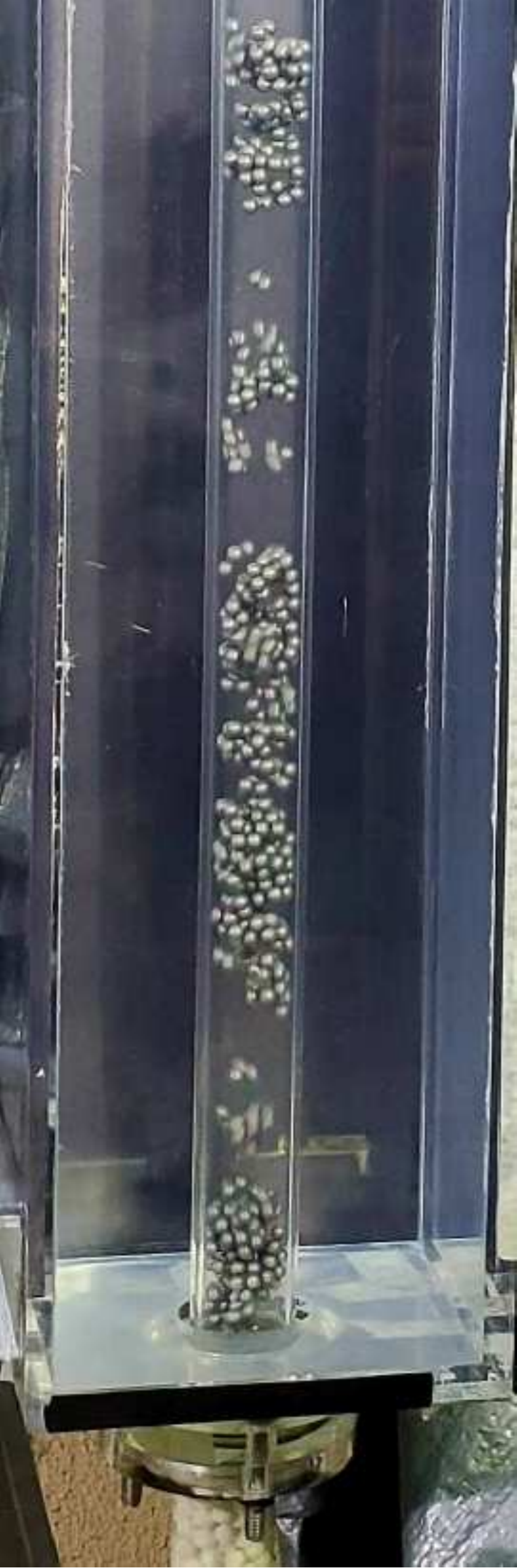}
	\caption{Fluidized bed occupying part of the test section.}
	\label{fig_test_section}
\end{figure}

The experimental setup consisted of a water tank with a heat exchanger, a centrifugal pump, a flow meter, a flow homogenizer, a 25.4-mm-ID vertical tube, and a return line, where water flowed in closed loop following the order just described. The vertical tube was 1.2 m long, of which the first 0.65 m corresponded to the test section, and it was made of transparent material (polymethyl methacrylate - PMMA). It was aligned vertically within $\pm 3^{\circ}$, and a visual box filled with water was placed around its test section to minimize optical distortions. Figure \ref{fig:1} shows the layout of the experimental setup and Fig. \ref{fig_test_section} shows a photograph of part of the test section.

The centrifugal pump had a maximum flow capacity of 4100 l/h of water, and upward flows at fixed flow rates were imposed by controlling the pump rotation. A flow homogenizer consisting of a 150-mm-long tube containing packed beads with $d$ = 6 mm between fine wire screens and placed upstream the test section assured uniform water flows at the tube inlet. The heat exchanger had a feedback mechanism that assured water temperatures within 25$^{\circ}$C $\pm$ 3$^{\circ}$C.

\begin{figure}[ht]
	\centering
	\includegraphics[width=0.3\columnwidth]{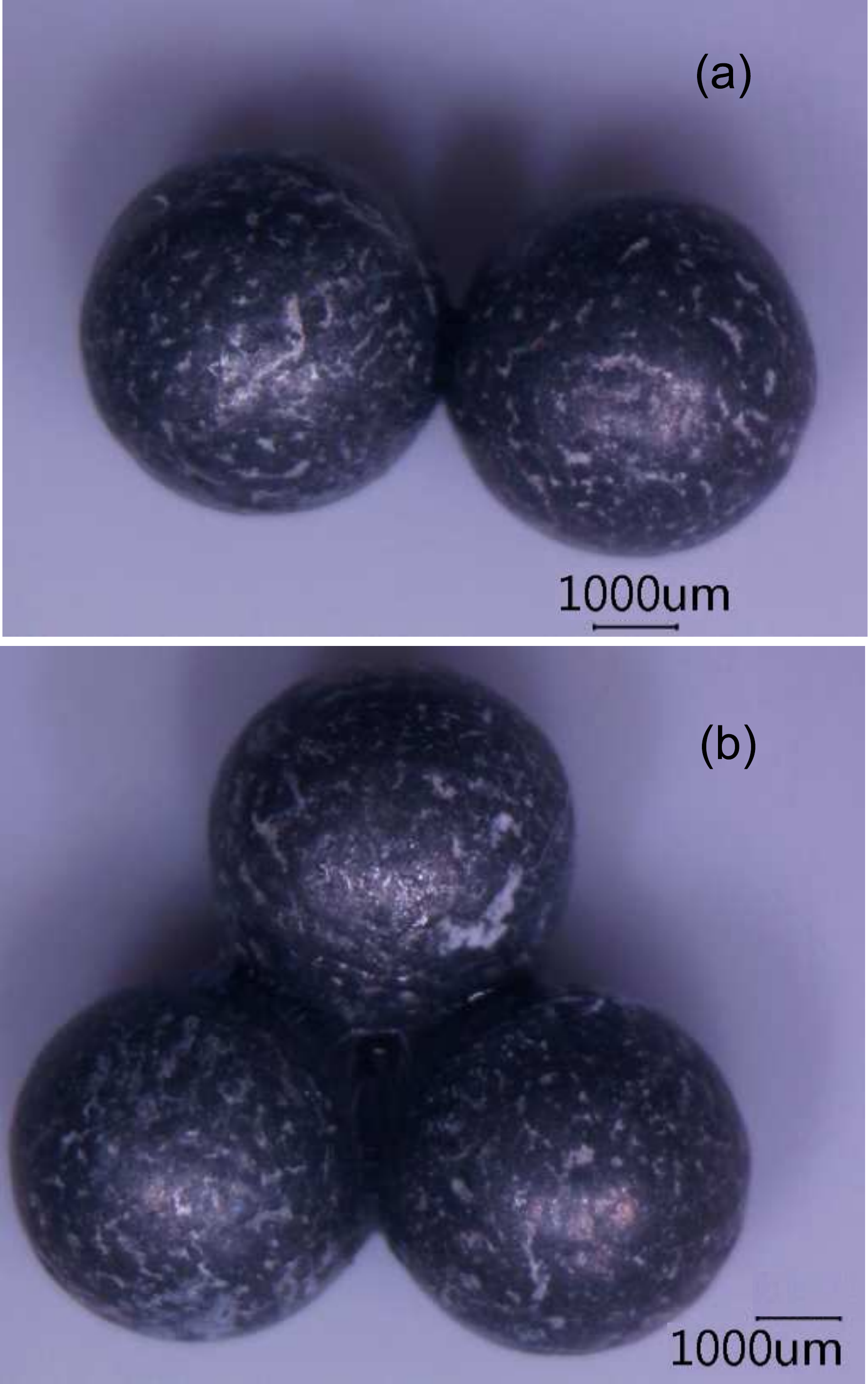}
	\caption{Microscopy images of (a) duos and (b) trios of bonded aluminum beads.}
	\label{fig_photo_bonded}
\end{figure}

Spherical beads of aluminum ($\rho_{p}$ = 2760 kg/m$^3$) and $d$ = 4.8 mm $\pm$ 0.03 mm were bonded with a small quantity of epoxy glue forming duos or trios as shown in Fig. \ref{fig_photo_bonded}. For the spherical beads, $D/d$ = 5.3, and the numbers of Stokes $St_t \,=\, v_t d \rho_p / (9\mu_f)$ and Reynolds $Re_t \,=\, \rho_f  v_t d / \mu_f$ based on the terminal velocity were 696 and 2269, respectively, where $v_t$ is the terminal velocity of one single sphere and $\mu_f$ is the dynamic viscosity of the fluid. The values of $St_t$ and $Re_t$ indicate that the used beads have considerable inertia with respect to water. The quantity of glue used for bonding was small, the mass of single spheres, duos and trios being of 0.15, 0.31 and 0.46 g, respectively. The mass of glue corresponds thus to 3\% and 2\% of the weight of duos and trios, respectively (see the supplementary material for microscopy images of duos and trios zoomed at the epoxy glue). 

The glued particles were, at the beginning of each test, left to settle by free fall in the test section, forming granular beds that consisted either of duos or trios and that were afterward fluidized by an upward water flow. In the present tests, we used a number of elements $N$ of 150 and 300 duos and 100 and 200 trios, and the cross-sectional average velocity of water $U$ varied within 0.137 and 0.192 m/s. Table \ref{table:table1} presents the bed heights $h_{if}$, particle fractions $\phi_0$ and water velocities $U_{if}$ at the inception of fluidization for each bed, and the settling velocity of individual spheres, $v_{s,i}$, computed with the Richardson--Zaki correlation for loose spheres, $v_{s,i} = v_t \left( 1-\phi_0 \right) ^{2.4}$. Values of $h_{if}$, $\phi_0$ and $U_{if}$ were determined experimentally by using image processing, where $\phi_0$ was computed as $h_{if}$ multiplied by the tube cross section and divided by the total volume of particles (number of particles multiplied by the particle volume).

\begin{table}[ht]
\caption{Type of element, number of elements $N$, bed height $h_{if}$, particle fraction $\phi_0$ and water velocity $U_{if}$ at the inception of fluidization, and settling velocity of individual spheres $v_{s,i}$}
\label{table:table1}
\centering
\begin{tabular}{c c c c c c}  
\hline\hline
Type & $N$ & $h_{if}$ & $\phi_0$  & $v_{s,i}$ & $U_{if}$\\
$\cdots$ & $\cdots$ & m & $\cdots$ & m/s & m/s\\ [0.5ex] 
\hline 
Duo & 150 & 0.075 & 0.46 & 0.110 & 0.119\\
Duo & 300 & 0.141 & 0.49 & 0.096 & 0.104\\
Trio & 100 & 0.078 & 0.44 & 0.118 & 0.129\\
Trio & 200 & 0.157 & 0.44 & 0.119 & 0.129\\
\hline
\hline 
\end{tabular}
\end{table}

Either a high-speed or a conventional camera was placed perpendicularly to the test section in order to acquire images of the beds. For measurements at the grain scale, we used a high-speed camera of complementary metal-oxide-semiconductor (CMOS) type with maximum resolution of 2560 px $\times$ 1600 px at 800 Hz, and we set its region of interest (ROI) to 2496 px $\times$ 216 px and frequency to 200 Hz. The field of view was of 362 mm $\times$ 31 mm, which corresponds to approximately 7 px/mm. For measurements over long timescales, we used a conventional camera of CMOS type with a maximum resolution of 1920 px $\times$ 1080 px at 60 Hz, and the ROI was set to 1910 px $\times$ 116 px at 60 Hz for a field of view of 418 mm $\times$ 25 mm, corresponding to approximately 5 px/mm. The cameras were branched to a computer system that controlled the cameras and pump rotation. We mounted lenses of $60$ mm focal distance and F2.8 maximum aperture on the cameras and made use of lamps of light-emitting diode (LED) branched to a continuous-current source.

\section{NUMERICAL SETUP}
\label{section_numerical_setup}

The use CFD-DEM is important since it gives access to the instantaneous positions of each element within the bed, the main drawback of our experiments being the bed opacity. Therefore, the trajectories of all individual elements and the network of contact forces are accessible from CFD-DEM outputs, while from experiments only two-dimensional projections of trajectories and bed structures can be measured.

\begin{sloppypar}
Three-dimensional computations were carried out with the open-source code CFDEM \cite{Goniva} (www.cfdem.com), that couples OpenFOAM, computing the fluid motion in an Eulerian frame using the finite volume method, to LIGGGHTS, computing the granular dynamics in a Lagrangian frame using DEM. For very narrow pipes, the mesh size must be small enough to capture the main characteristics of the flow, while the particle diameter is relatively large, making the resolved formulation more appropriate. However, resolved coupling demands very refined meshes around each particle, with consequent high computational costs. One way to circumvent this problem, and that was employed in our simulations, is by using an unresolved approach together with a CFDEM model called \textit{big particle}, in which the region of influence of each particle is artificially increased, being compensated by a fictitious porosity \cite{Kloss2,Mondal}.
\end{sloppypar}

Basically, we set the CFD-DEM code to solve Eqs. \ref{Fp} to \ref{qdm}. The dynamics of each solid particle is computed by using the linear and angular momentum equations, given by Eqs. \ref{Fp} and \ref{Tp}, respectively:

\begin{equation}
m_{p}\frac{d\vec{u}_{p}}{dt}= \vec{F}_{D} + \vec{F}_{stress} + \vec{F}_{am} + \vec{F}_g + \vec{F}_{c} \,,
\label{Fp}
\end{equation}

\begin{equation}
I_{p}\frac{d\vec{\omega}_{p}}{dt}=\vec{T}_{c} \,,
\label{Tp}
\end{equation}

\noindent where, for each solid particle, $m_{p}$ is the mass, $\vec{u}_{p}$ is the velocity, $I_{p}$ is the moment of inertia, $\vec{\omega}_{p}$ is the angular velocity, $\vec{F}_{c}$ is the resultant of contact forces between solids, $\vec{T}_{c}$ is the resultant of contact torques between solids, $\vec{F}_{D}$ is the drag force caused by the fluid on particles, $\vec{F}_g$ is the gravitational force, $\vec{F}_{stress}$ = $V_{p}\left[ -\nabla P + \nabla \cdot\vec{\vec{\tau}} \right]$ is the force caused by fluid stresses and $\vec{F}_{am}$ is the added mass force, $V_{p}$ being the volume of each solid particle, $P$ the fluid pressure and $\vec{\vec{\tau}}$ the deviatoric stress tensor of the fluid. For the fluid, conservation of mass and momentum are computed with Eqs. \ref{mass} and \ref{qdm}, respectively:

\begin{equation}
{\frac{\partial{\rho_{f}\varepsilon_{f}}}{\partial{t}}+\nabla\cdot(\rho_{f}\varepsilon_{f}\vec{u}_{f})=0} \,,
\label{mass}
\end{equation}

\begin{equation}
\frac{\partial{\rho_{f}\varepsilon_{f}\vec{u}_{f}}}{\partial{t}} + \nabla \cdot (\rho_{f}\varepsilon_{f}\vec{u}_{f}\vec{u}_{f}) = -\varepsilon_{f}\nabla P + \varepsilon_{f}\nabla\cdot \vec{\vec{\tau}}_{f} - \frac{\vec{F}_{D}}{V_{cell}} \,,
\label{qdm}
\end{equation}

\noindent where $\vec{u}_{f}$ and $\varepsilon_{f}$ are, respectively, the mean velocity and volume fraction of the fluid phase, and $V_{cell}$ is the volume of the considered cell.

\begin{figure}[ht]
	\begin{minipage}{0.49\linewidth}
		\begin{tabular}{c}
			\includegraphics[width=0.31\linewidth]{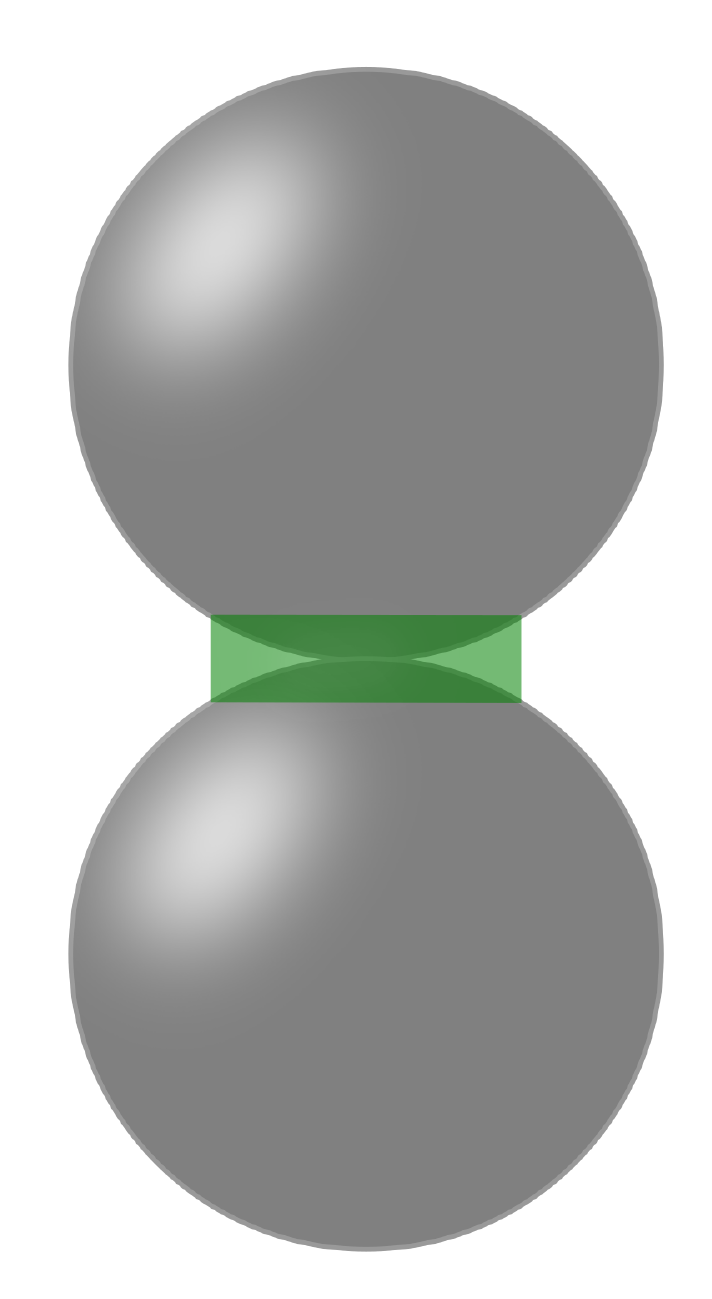}\\
			(a)
		\end{tabular}
	\end{minipage}
	\hfill
	\begin{minipage}{0.49\linewidth}
		\begin{tabular}{c}
			\includegraphics[width=0.55\linewidth]{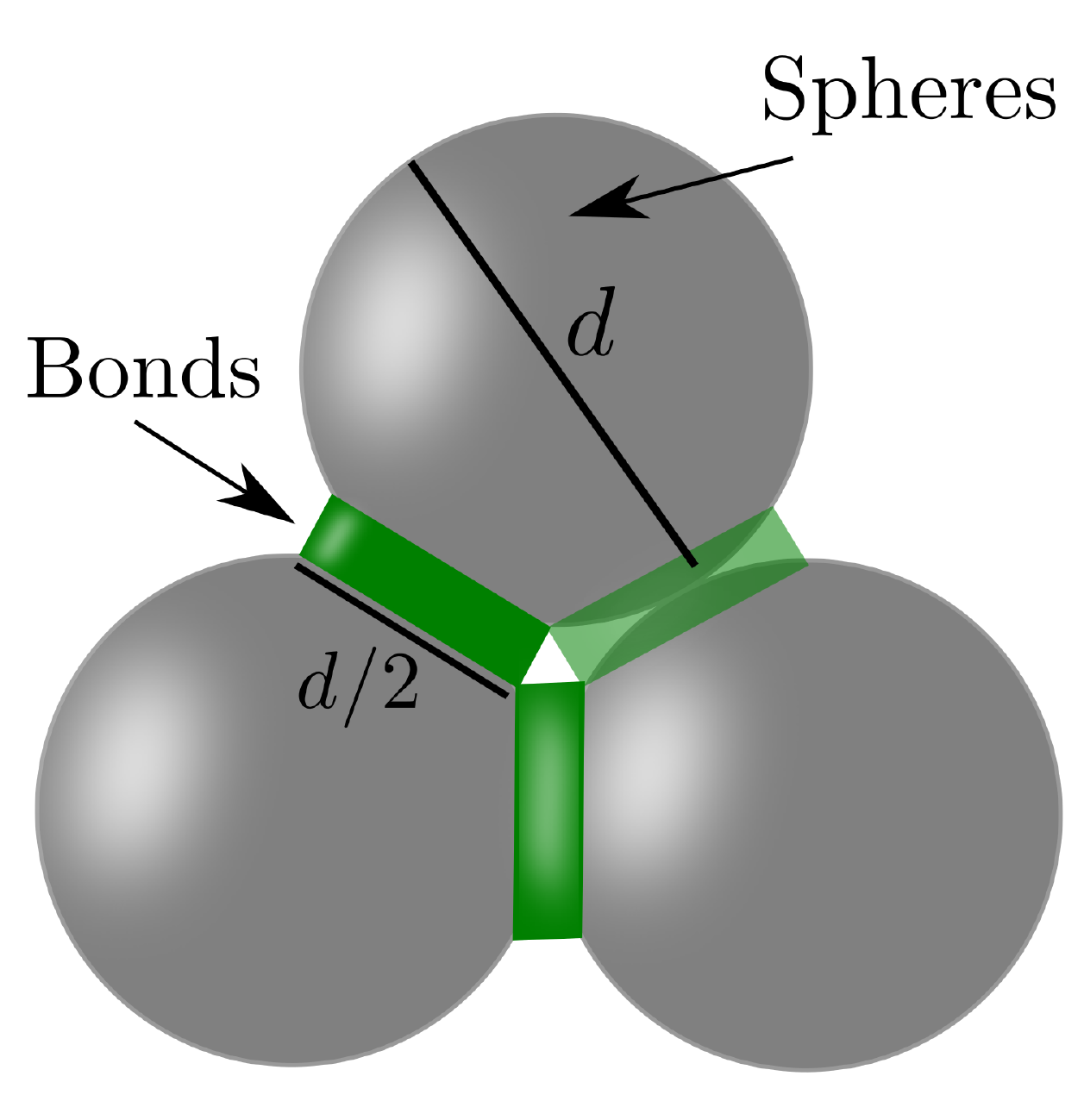}\\
			(b)
		\end{tabular}
	\end{minipage}
	\hfill
	\caption{Numerically built (a) duos and (b) trios.}
	\label{fig:bonds_num}
\end{figure}

We consider SLFBs consisting of bonded spheres forming duos or trios, under the same conditions as in the experiments. Duos and trios were implemented numerically by putting into permanent contact spheres that do not overlap each other, with bonds that have a diameter equal to half of that of spheres and are considered to be solid, filling the internal space in a similar manner as the elements used in experiments (see the supplementary material for microscopy images of duos and trios zoomed at the epoxy glue). Figures \ref{fig:bonds_num}a and \ref{fig:bonds_num}b show the numerical duos and trios, respectively. The spheres are considered to be of aluminum and their properties are summarized in Tab. \ref{tab_properties}, together with that of water.

\begin{table}[ht]
\centering
\caption{Physical properties of the fluid and solid particles.}
\begin{tabular}{ll}
\hline
Sphere diameter $d$ (mm)                         & 4.8   \\
Sphere density $\rho_p$ (kg/m$^{3}$)                            & 2760  \\
Young's modulus $E$ (GPa)                          & 71    \\
Poisson ratio $\sigma$                               & 0.34  \\
Restitution coefficient $e$                          & 0.5   \\
Particle-particle friction coefficient $\mu_{fr, p}$ & 0.5   \\
Particle-wall friction coefficient $\mu_{fr, w}$     & 0.9   \\
Liquid density $\rho_{f}$ (kg/m$^{3}$)               & 1000  \\
Liquid viscosity $\mu_{f}$ (Pa.s)                  & 0.001 \\ \hline
\end{tabular}
\label{tab_properties}
\end{table}
 
\begin{figure}[ht]
	\centering
	\includegraphics[width=0.60\columnwidth]{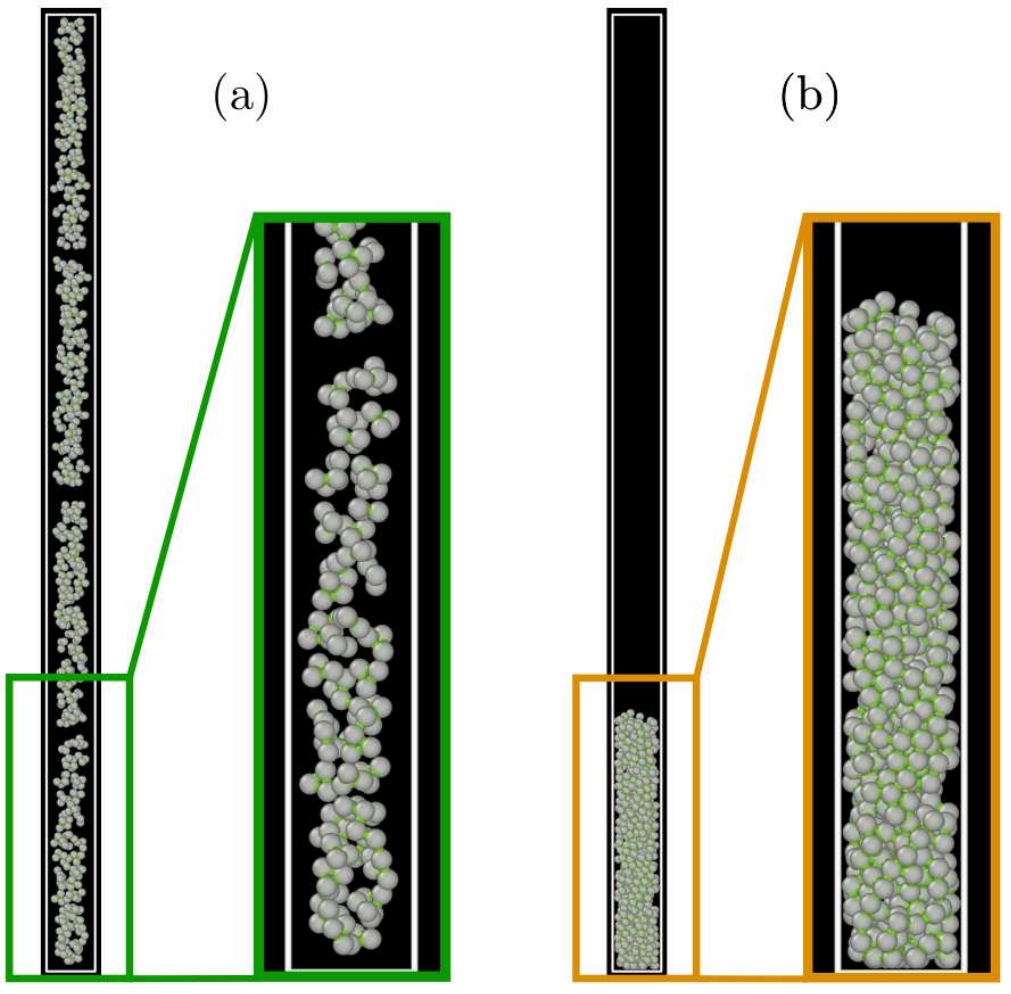}
	\caption{Initial conditions for simulations of fluidized beds: (a) settling of randomly distributed elements; (b) relaxed state after settling.}
	\label{fig_init_conditions}
\end{figure}

The problem is initialized with a random distribution and orientation of elements throughout a 25.4-mm-ID and 0.55-m-long pipe. At first, the elements are allowed to fall freely in still water until a steady state is reached. Afterward, an ascendant water flow begins and increases until reaching the desired velocity after 1 s. An initial particle distribution as well as a steady settled bed are presented in Fig. \ref{fig_init_conditions}.

For the fluid, a no-slip boundary condition was applied at the tube wall, fixed velocity and zero pressure gradient at the tube inlet, and zero velocity gradient and fixed pressure at the tube outlet. At the inlet, the initial velocity is interpolated from zero to the desired value over the first second of the simulation. The computational mesh consisted of 36250 hexaedral cells, with a maximum cell volume of approximately 1.15 $\times$ 10$^{-8}$ m$^{3}$ (see the supplementary material for a figure showing the computational domain).

Fluid equations are solved using Pressure-Implicit with Splitting of Operators (PISO) with two main pressure corrections and one non-orthogonal flux correction to account for the slightly non-orthogonal cells generated by the cylindrical geometry. The DEM time step was set to 1 $\times$ 10$^{-6}$ s, which is below 10\% of Rayleigh and Hertz time scales \cite{Derakhshani}, and CFD time step was set to 5 $\times$ 10$^{-5}$ s, corresponding to 50 DEM time steps.

\section{RESULTS AND DISCUSSIONS}

\subsection{Macroscopic observations}
 
\begin{figure}[ht]
	\begin{center}
	\begin{tabular}{c}
	\includegraphics[width=0.8\columnwidth]{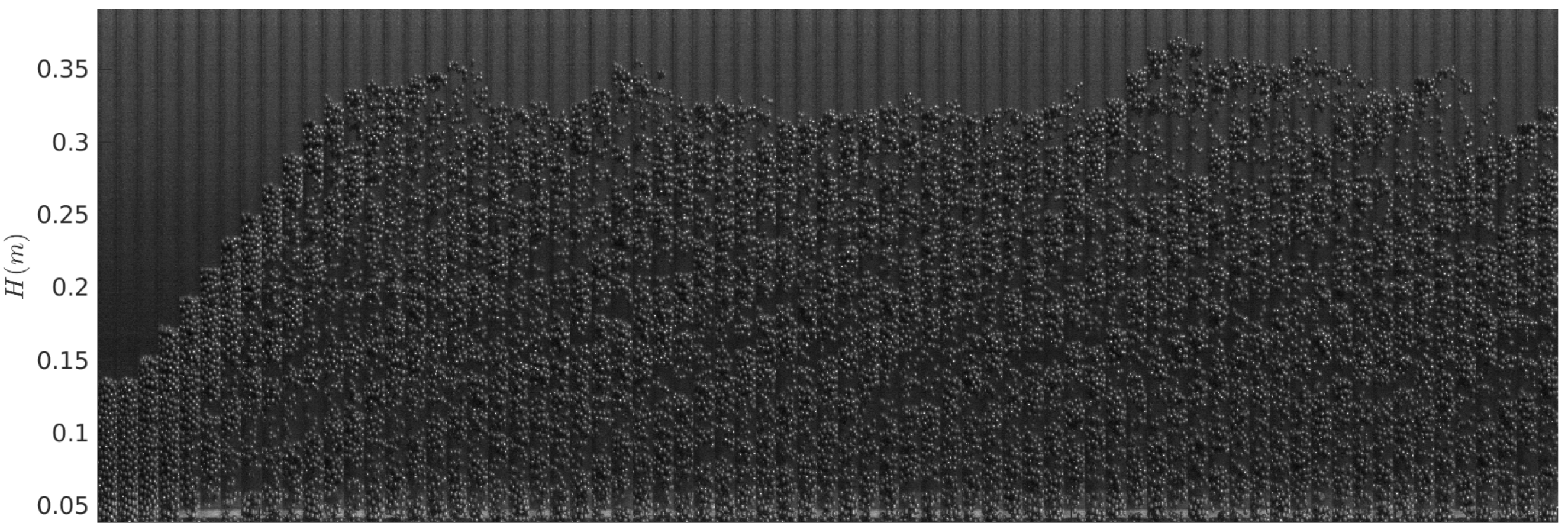}\\
	(a)\\
	\includegraphics[width=0.8\columnwidth]{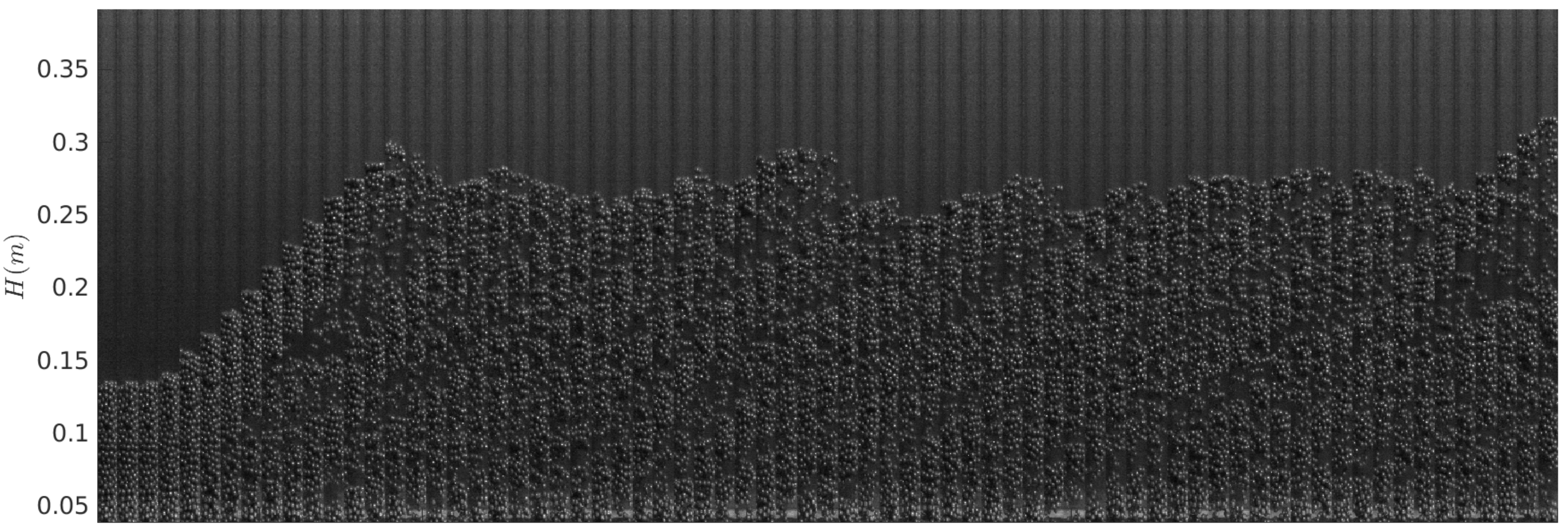}\\
	(b)\\
	\includegraphics[width=0.8\columnwidth]{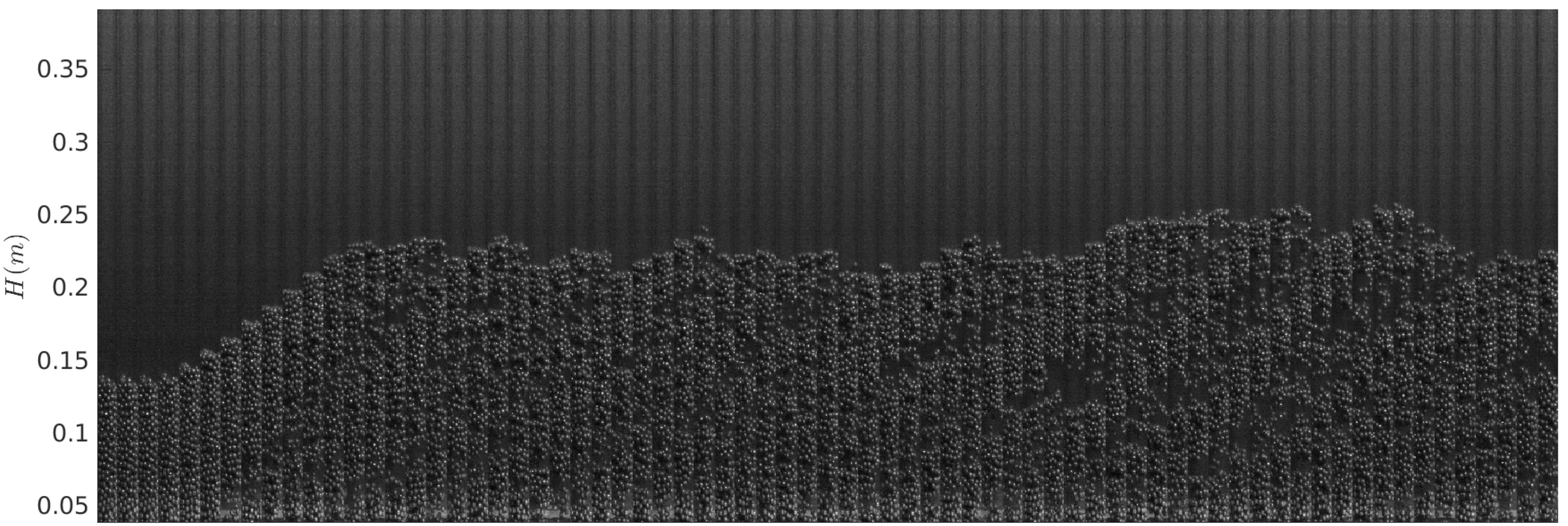}\\
	(c)
	\end{tabular} 
\end{center}
	\caption{Snapshots of particle positions, from experiments, for beds consisting of 300 duos, placed side by side. (a) $U$ = 0.192 m/s; (b) $U$ = 0.164 m/s; (c) $U$ = 0.137 m/s (Multimedia view, where the reproduction rate of the movie is 0.1 times the real time). The total time is 14 s and time between frames 0.2 s.}
	\label{fig:snapshots_duos_exp}
\end{figure}

\begin{figure}[ht]
	\begin{center}
	\begin{tabular}{c}
	\includegraphics[width=0.8\columnwidth]{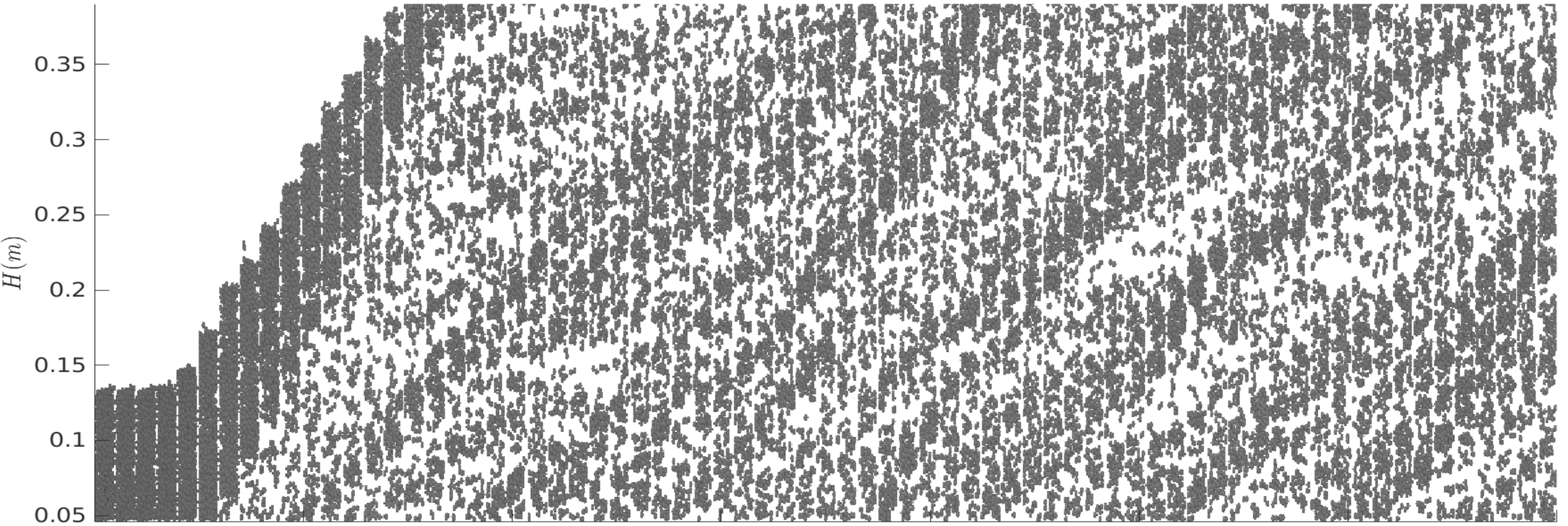}\\
	(a)\\
	\includegraphics[width=0.8\columnwidth]{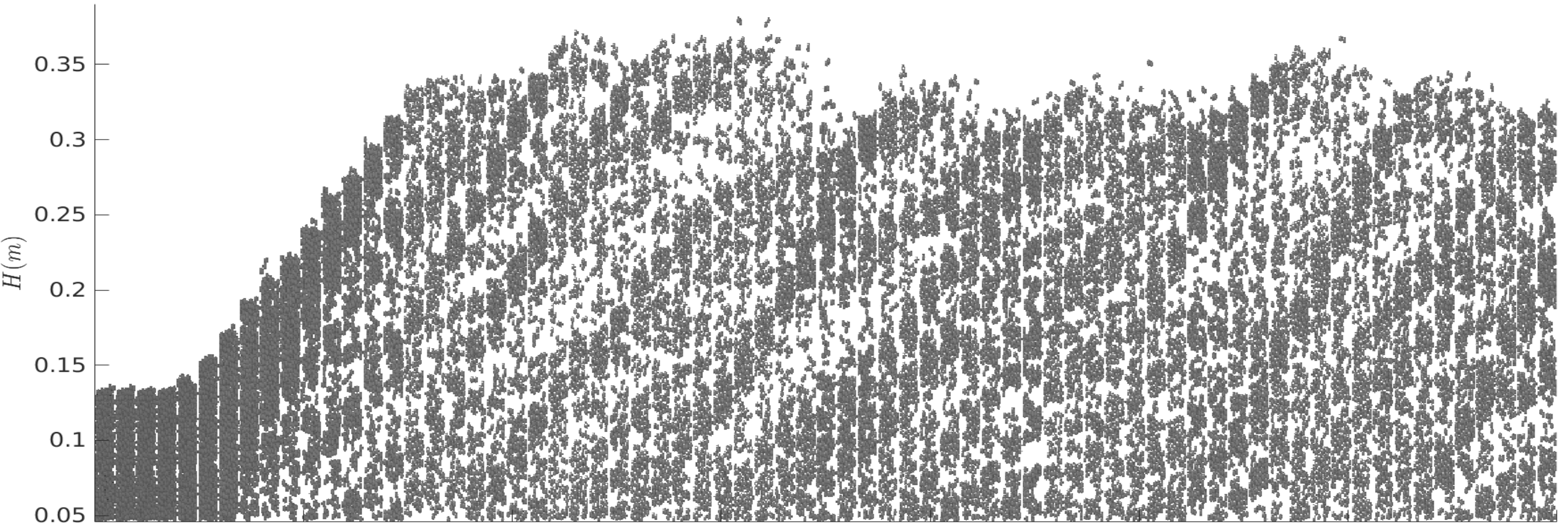}\\
	(b)\\
	\includegraphics[width=0.8\columnwidth]{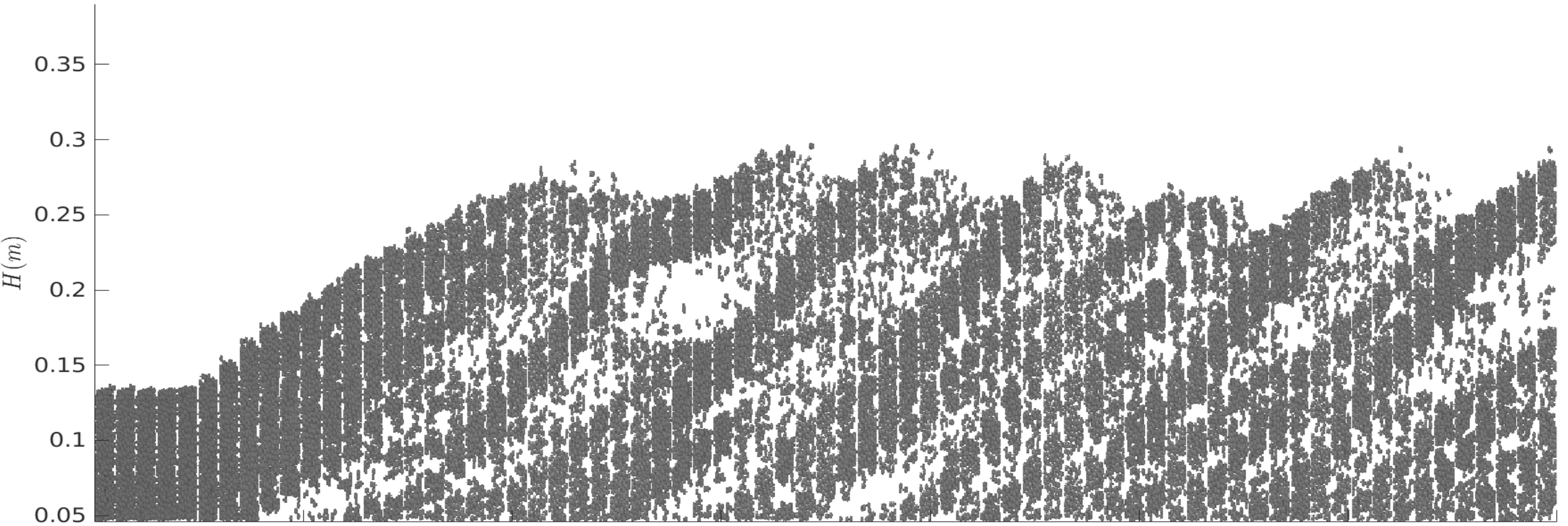}\\
	(c)
	\end{tabular} 
\end{center}
	\caption{Snapshots of particle positions, from numerical simulations, for beds consisting of 300 duos, placed side by side. (a) $U$ = 0.192 m/s; (b) $U$ = 0.164 m/s; (c) $U$ = 0.137 m/s (Multimedia view, where the reproduction rate of the movie is 0.1 times the real time). The total time is 14 s and time between frames 0.2 s.}
	\label{fig:snapshots_duos_num}
\end{figure}

\begin{figure}[ht]
	\begin{center}
	\begin{tabular}{c}
	\includegraphics[width=0.8\columnwidth]{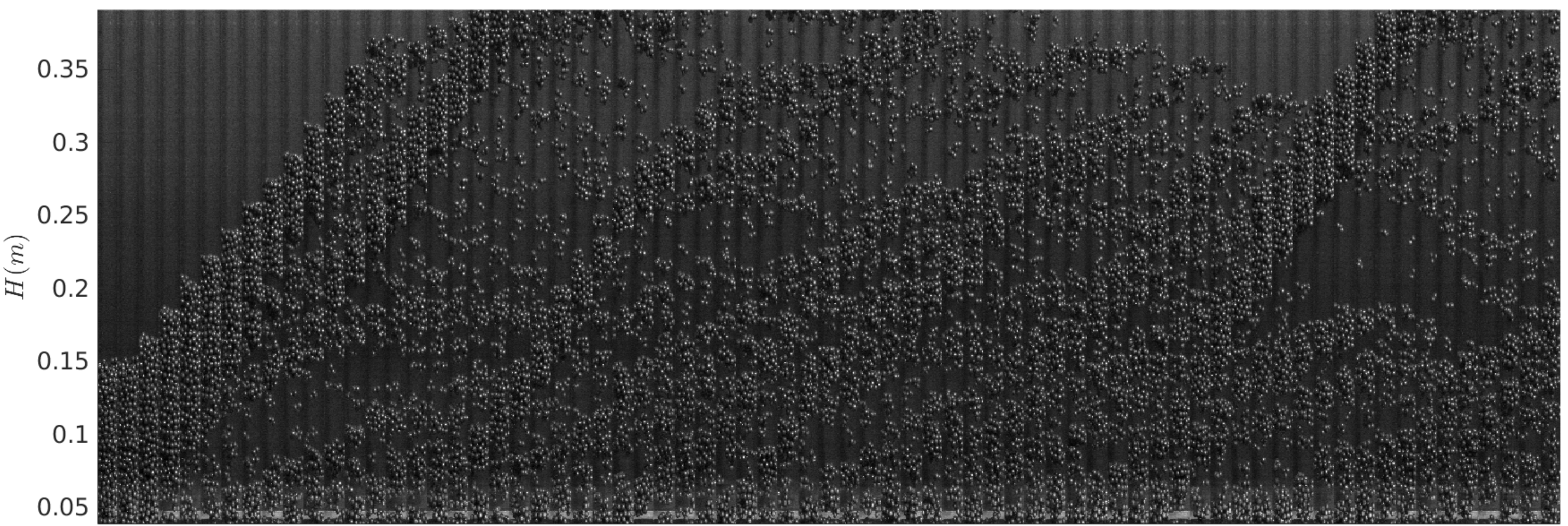}\\
	(a)\\
	\includegraphics[width=0.8\columnwidth]{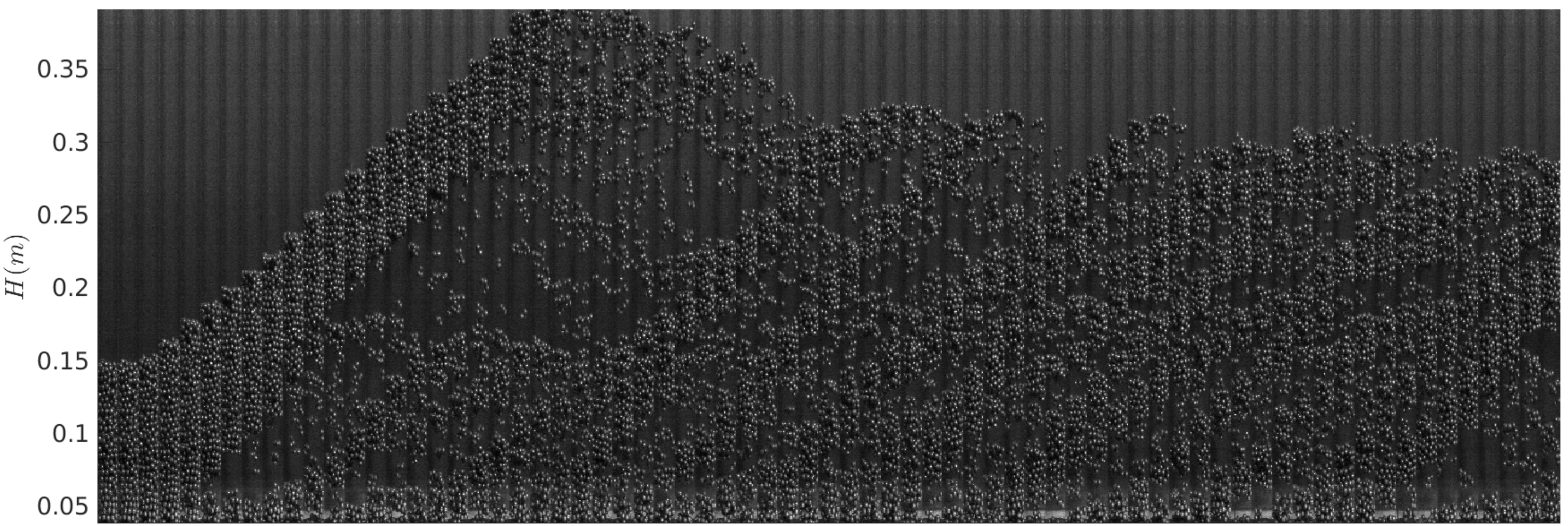}\\
	(b)\\
	\includegraphics[width=0.8\columnwidth]{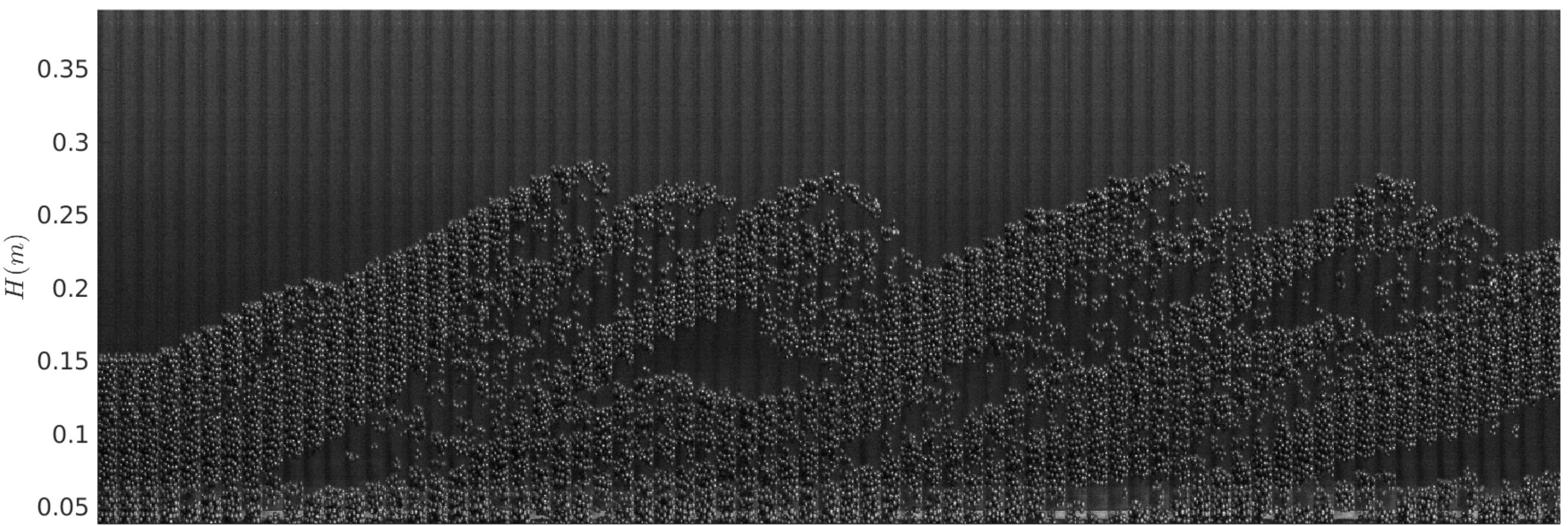}\\
	(c)
	\end{tabular} 
\end{center}
	\caption{Snapshots of particle positions, from experiments, for beds consisting of 200 trios, placed side by side. (a) $U$ = 0.192 m/s; (b) $U$ = 0.164 m/s; (c) $U$ = 0.137 m/s (Multimedia view, where the reproduction rate of the movie is 0.1 times the real time). The total time is 14 s and time between frames 0.2 s.}
	\label{fig:snapshots_trios_exp}
\end{figure}

\begin{figure}[ht]
	\begin{center}
	\begin{tabular}{c}
	\includegraphics[width=0.8\columnwidth]{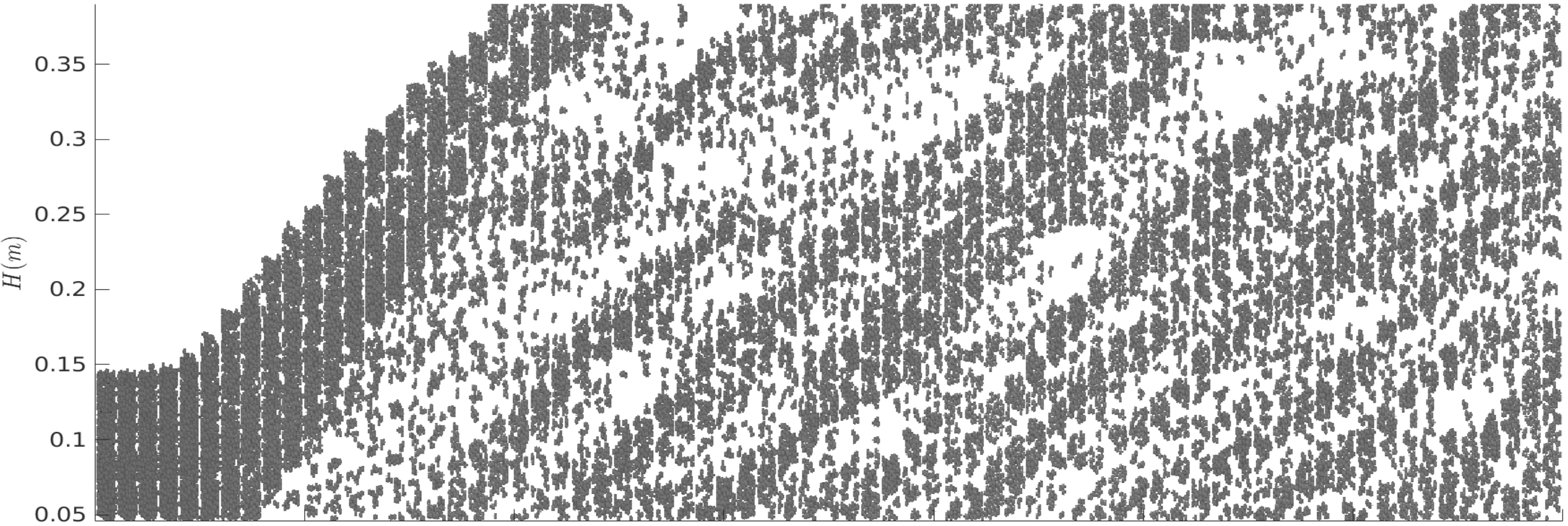}\\
	(a)\\
	\includegraphics[width=0.8\columnwidth]{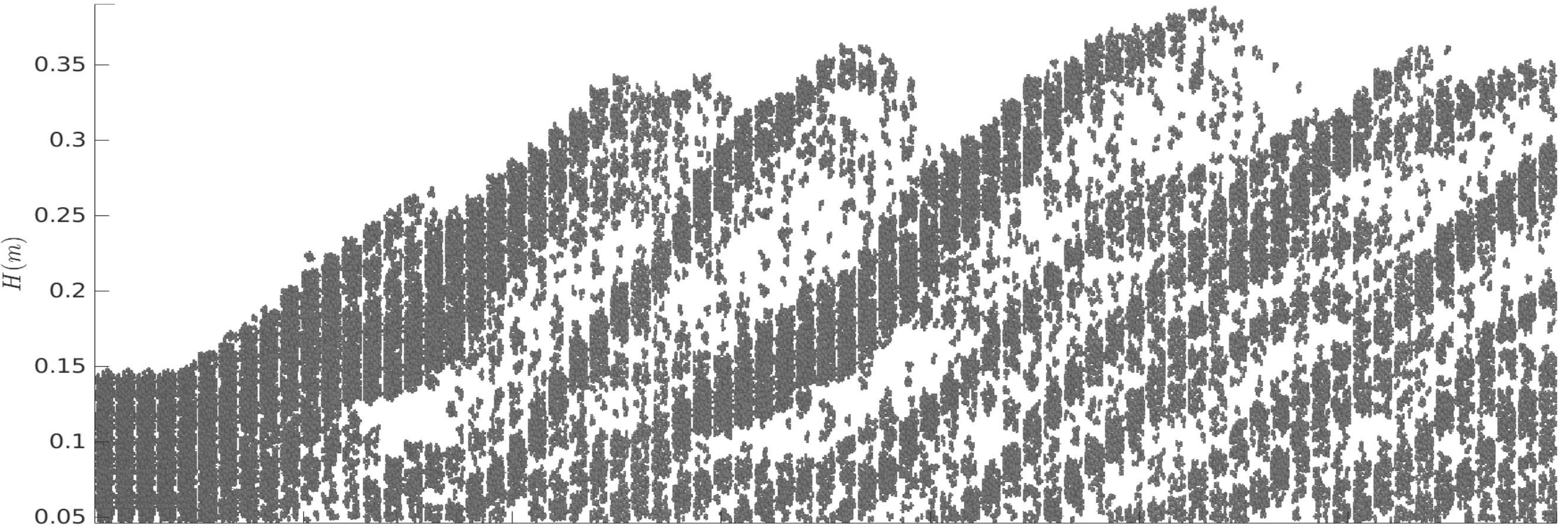}\\
	(b)\\
	\includegraphics[width=0.8\columnwidth]{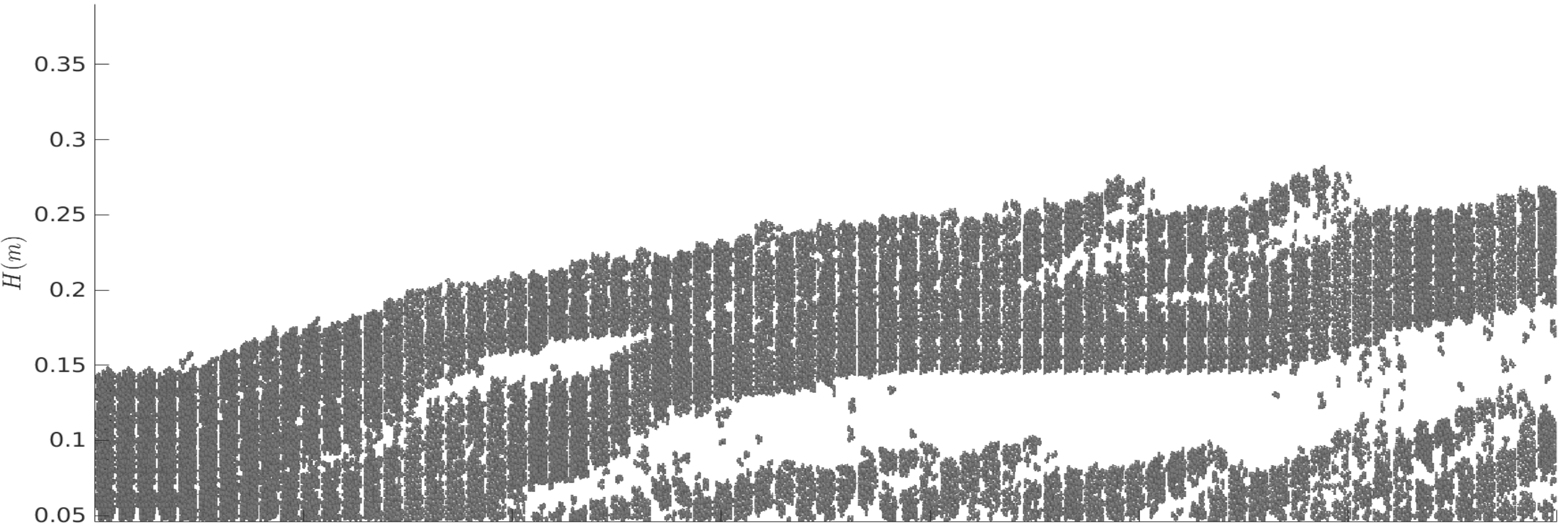}\\
	(c)
	\end{tabular} 
\end{center}
	\caption{Snapshots of particle positions, from numerical simulations, for beds consisting of 200 trios, placed side by side. (a) $U$ = 0.192 m/s; (b) $U$ = 0.164 m/s; (c) $U$ = 0.137 m/s (Multimedia view, where the reproduction rate of the movie is 0.1 times the real time). The total time is 14 s and time between frames 0.2 s.}
	\label{fig:snapshots_trios_num}
\end{figure}

We observed the development of bed structures in the form of plugs, such as those for loose grains \cite{Cunez,Cunez2}, but with lower mobility. In the case of trios, plugs were frequently blocked at upper positions in the tube, and jamming occurred at much higher water velocities when compared to loose grains \cite{Cunez3}. In order to show the macroscopic structure of the bed, we placed side by side snapshots of particle positions, for both experiments and simulations, and present them, respectively, in Figs. \ref{fig:snapshots_duos_exp} and \ref{fig:snapshots_duos_num} for 300 duos and Figs. \ref{fig:snapshots_trios_exp} and \ref{fig:snapshots_trios_num} for 200 trios. In Figs. \ref{fig:snapshots_duos_exp} to \ref{fig:snapshots_trios_num}, total duration is 14 s and time between frames 0.2 s, and Figs. \ref{fig:snapshots_duos_exp}c, \ref{fig:snapshots_duos_num}c, \ref{fig:snapshots_trios_exp}c and \ref{fig:snapshots_trios_num}c are of \textit{Multimedia view} type, presenting movies with reproduction rate of 0.1 times the real time in the online version (see the supplementary material or Mendeley Data \cite{Supplemental2} for snapshots of the other cases). For comparison with what we present next, snapshots of beds consisting of loose spheres and a table with lengths, celerities and frequencies for those beds are available in the supplementary material.

From Figs. \ref{fig:snapshots_duos_exp} to \ref{fig:snapshots_trios_num}, we note that for lower velocities a transient occurs at the beginning of test runs, in which an upward displacement of most of the bed takes place in the form of a single granular plug, while for the largest velocity the bed breaks into smaller structures from the beginning of tests, though large plugs can be observed afterward. The initial transient and its dependence on the water velocity are more pronounced in the case of trios. After the initial transient, we observe the passage of granular plugs and void regions, with the top of the bed oscillating between maximum and minimum values. By reducing the water velocity, going from Fig. \ref{fig:snapshots_duos_exp}a to \ref{fig:snapshots_duos_exp}c and from Fig. \ref{fig:snapshots_trios_exp}a to \ref{fig:snapshots_trios_exp}c, we observe that plugs become longer and more frequent in the bed.

\begin{figure}[ht]
	\begin{center}
	\begin{tabular}{c}
	\includegraphics[width=0.8\columnwidth]{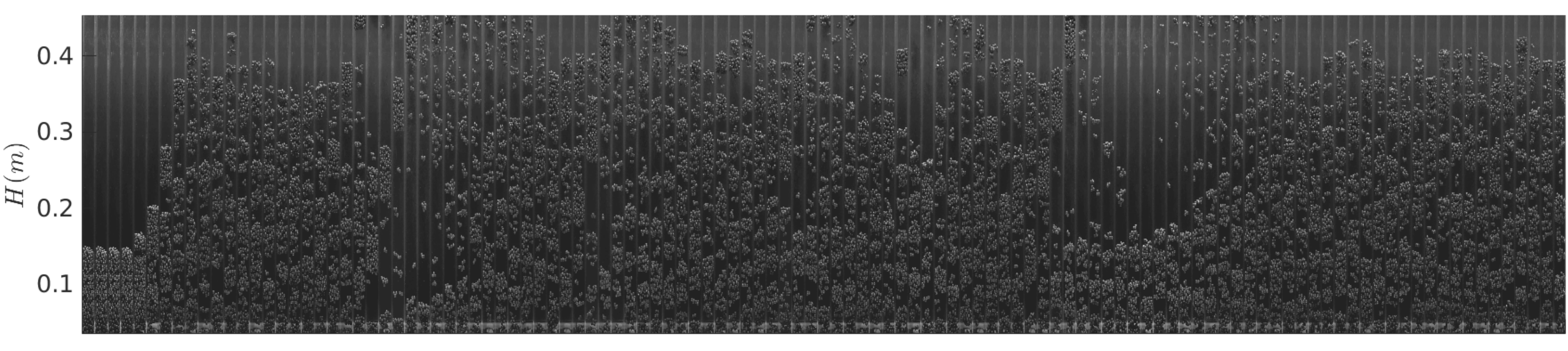}\\
	(a)\\
	\includegraphics[width=0.8\columnwidth]{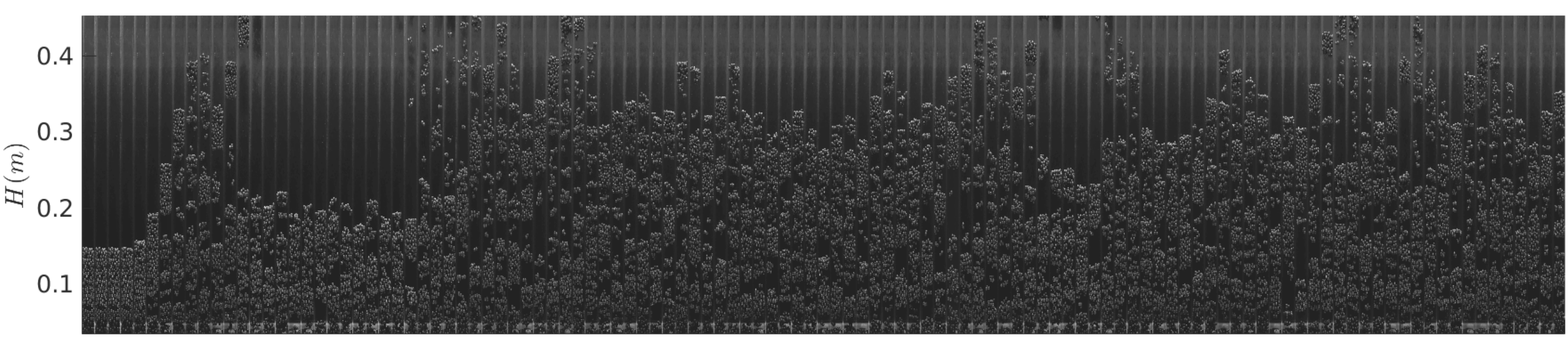}\\
	(b)\\
	\includegraphics[width=0.8\columnwidth]{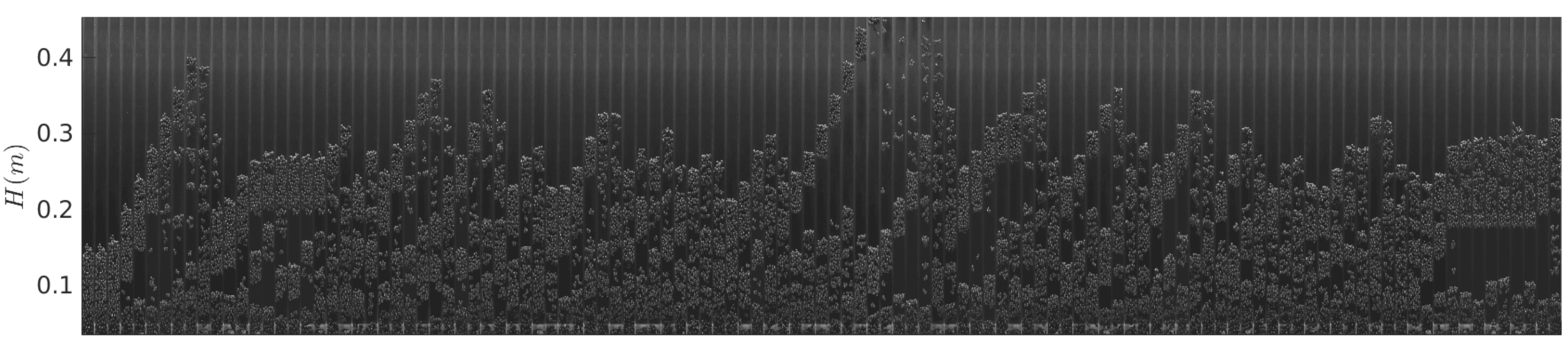}\\
	(c)\\
	\includegraphics[width=0.8\columnwidth]{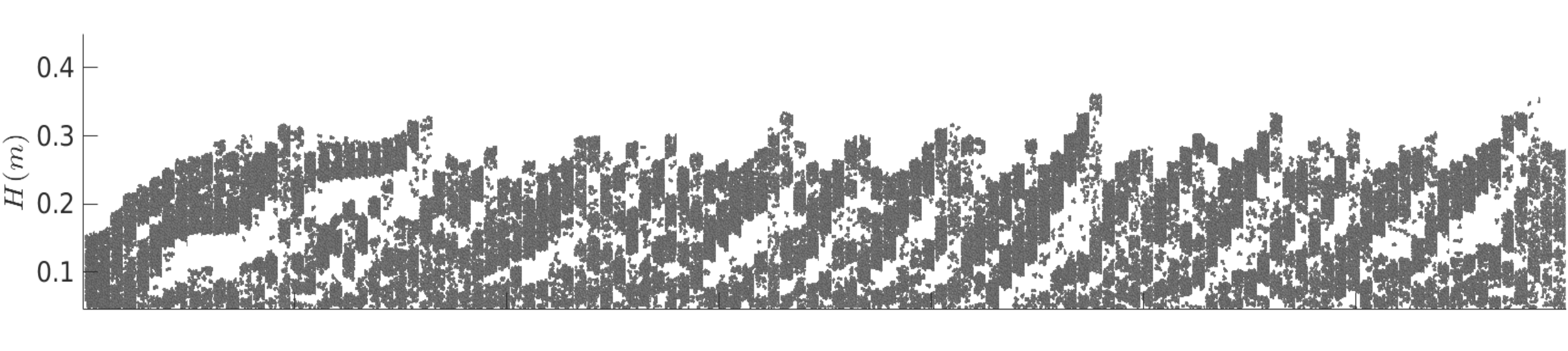}\\
	(d)
	\end{tabular} 
\end{center}
	\caption{Snapshots placed side by side of particle positions for beds consisting of 200 trios. (a) $U$ = 0.192 m/s; (b) $U$ = 0.164 m/s; (c) and (d) $U$ = 0.137 m/s. (a) to (c) correspond to experiments and (d) to numerical simulations. The total time is 105 s and time between frames 1 s.}
	\label{fig:snapshots_trio}
\end{figure}

Fig. \ref{fig:snapshots_trio} shows snapshots of 200 trios over longer times, in order to investigate if the bed structure changes along time. Figures \ref{fig:snapshots_trio}a to \ref{fig:snapshots_trio}c correspond to experiments and Fig. \ref{fig:snapshots_trio}d to numerical simulations, the total time being 105 s and time between frames 1 s. These snapshots are representative of other tests with trios over long times, and tests with duos did not show changes over long durations; therefore, the latter are not presented here. The same observations done for Figs. \ref{fig:snapshots_trios_exp} and \ref{fig:snapshots_trios_num} can be made here, until, for the smallest velocity (Fig. \ref{fig:snapshots_trio}c), a granular plug is seen to be blocked in the middle of the tube by the end of the test. For some tests, grains within the clogging plug revealed to be jammed, as we investigate in Subsection \ref{subsection_jamming}.

Initially, we processed the acquired images to measure the general bed structure, namely the average bed height $H$, average upward celerity of its top $c_{up}$, average downward celerity of its top $c_{down}$, and average plug length $\lambda$. Table \ref{table:table3} presents average values from experiments (Exp) and numerical simulations (Num), for each tested condition, of $H$, $c_{up}$, $c_{down}$ and $\lambda$, normalized by $D$ and $U_{if}$. Values in dimensional form and their respective standard deviations are available in the supplementary material and on Mendeley Data \cite{Supplemental2}.

\begin{table}[h!]
\caption{Source, type of element, number of elements $N$, cross-sectional water velocity $U$, average bed height $H$, average upward celerity of its top $c_{up}$, average downward celerity of its top $c_{down}$, and the average length of plugs $\lambda$. Values are normalized by $D$ and $U_{if}$.}
\label{table:table3}
\centering
\begin{tabular}{c c c c c c c c}  
\hline\hline
Source & Type & $N$ & $U/U_{if}$ & $H/D$ & $c_{up}/U_{if}$ & $c_{down}/U_{if}$ & $\lambda/D$\\
\hline 
Exp & Duo & 150 & 1.2 & 4.4 & 0.23 & -0.27 & 1.3\\
Exp & Duo & 150 & 1.4 & 5.5 & 0.30 & -0.33 & 1.2\\
Exp & Duo & 150 & 1.6 & 6.7 & 0.38 & -0.35 & 1.0\\
Exp & Duo & 300 & 1.3 & 8.8 & 0.32 & -0.38 & 1.7\\
Exp & Duo & 300 & 1.6 & 10.4 & 0.43 & -0.38 & 1.5\\
Exp & Duo & 300 & 1.8 & 12.8 & 0.47 & -0.43 & 1.1\\
Exp & Trio & 100 & 1.1 & 4.6 & 0.18 & -0.30 & 1.3\\
Exp & Trio & 100 & 1.3 & 5.6 & 0.27 & -0.31 & 1.0\\
Exp & Trio & 100 & 1.5 & 7.2 & 0.38 & -0.44 & 0.8\\
Exp & Trio & 200 & 1.1 & 10.0 & 0.25 & -0.40 & 1.4\\
Exp & Trio & 200 & 1.3 & 12.1 & 0.36 & -0.38 & 1.1\\
Exp & Trio & 200 & 1.5 & 13.9 & 0.49 & -0.43 & 0.9\\
Num & Duo & 150 & 1.2 & 4.9 & 0.27 & -0.47 & 1.6\\
Num & Duo & 150 & 1.4 & 6.2 & 0.36 & -0.40 & 1.2\\
Num & Duo & 150 & 1.6 & 8.6 & 0.45 & -0.44 & 0.9\\
Num & Duo & 300 & 1.3 & 10.1 & 0.39 & -0.41 & 1.5\\
Num & Duo & 300 & 1.6 & 12.6 & 0.55 & -0.50 & 1.2\\
Num & Duo & 300 & 1.8 & 15.9 & 0.64 & -0.59 & 1.1\\
Num & Trio & 100 & 1.1 & 4.6 & 0.16 & -0.41 & 1.4\\
Num & Trio & 100 & 1.3 & 5.9 & 0.32 & -0.52 & 1.0\\
Num & Trio & 100 & 1.5 & 7.9 & 0.46 & -0.68 & 1.1\\
Num & Trio & 200 & 1.1 & 9.3 & 0.15 & -0.26 & 1.8\\
Num & Trio & 200 & 1.3 & 12.2 & 0.38 & -0.88 & 1.4\\
Num & Trio & 200 & 1.5 & 16.2 & 0.45 & -0.48 & 1.0\\
\hline
\hline 
\end{tabular}
\end{table}

From Tab. \ref{table:table3}, together with Figs. \ref{fig:snapshots_duos_exp} to \ref{fig:snapshots_trio}, we note that, with increasing the water flow, the bed expands significantly, with oscillations of its top around increasing average values. The average height for trios is higher than that for duos for cases with 600 bonded spheres, and the reason seems related to the great drag coefficients $C_d$ experienced by trios when compared to duos. In order to verify that, we performed resolved simulations using the immersed boundary (IB) method for duos and trios in different orientations, and computed the respective values of $C_d$. The results, presented in the supplementary material, show that, indeed, drag coefficients are greater for trios. The bed top oscillates at a frequency of approximately 1.5 Hz for duos and trios, as well as for single spheres, varying slightly with the water flow as shown in the supplementary material. The top oscillation occurs as granular plugs and void regions pass by the top of the bed. We observe also that plugs with length $\lambda$ of the order of the tube diameter appear for both duos and trios, and that $\lambda$ decreases with increasing the water velocity. It is important to note that dispersions around mean values are relatively high, since plugs consist of rather large elements whose addition or removal by units change significantly instantaneous values (see the supplementary material for standard deviations of all experiments and simulations). That considered, results from numerical simulations are reasonably in accordance with experiments, which is also illustrated by Figs. \ref{fig:snapshots_duos_exp} to \ref{fig:snapshots_trio}.

\begin{figure}[ht]
	\begin{minipage}{0.49\linewidth}
		\begin{tabular}{c}
			\includegraphics[width=0.85\linewidth]{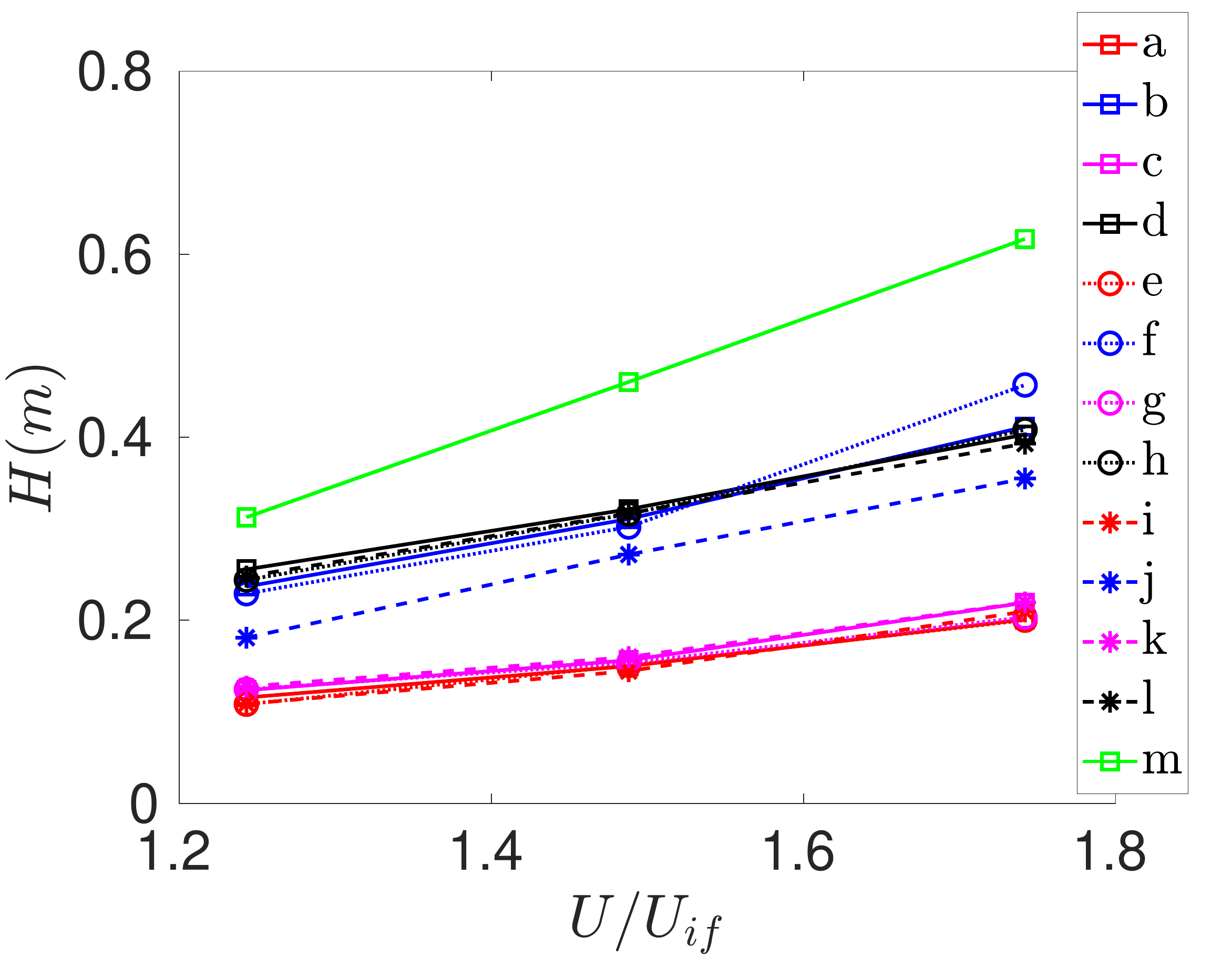}\\
			(a)
		\end{tabular}
	\end{minipage}
	\hfill
	\begin{minipage}{0.49\linewidth}
		\begin{tabular}{c}
			\includegraphics[width=0.85\linewidth]{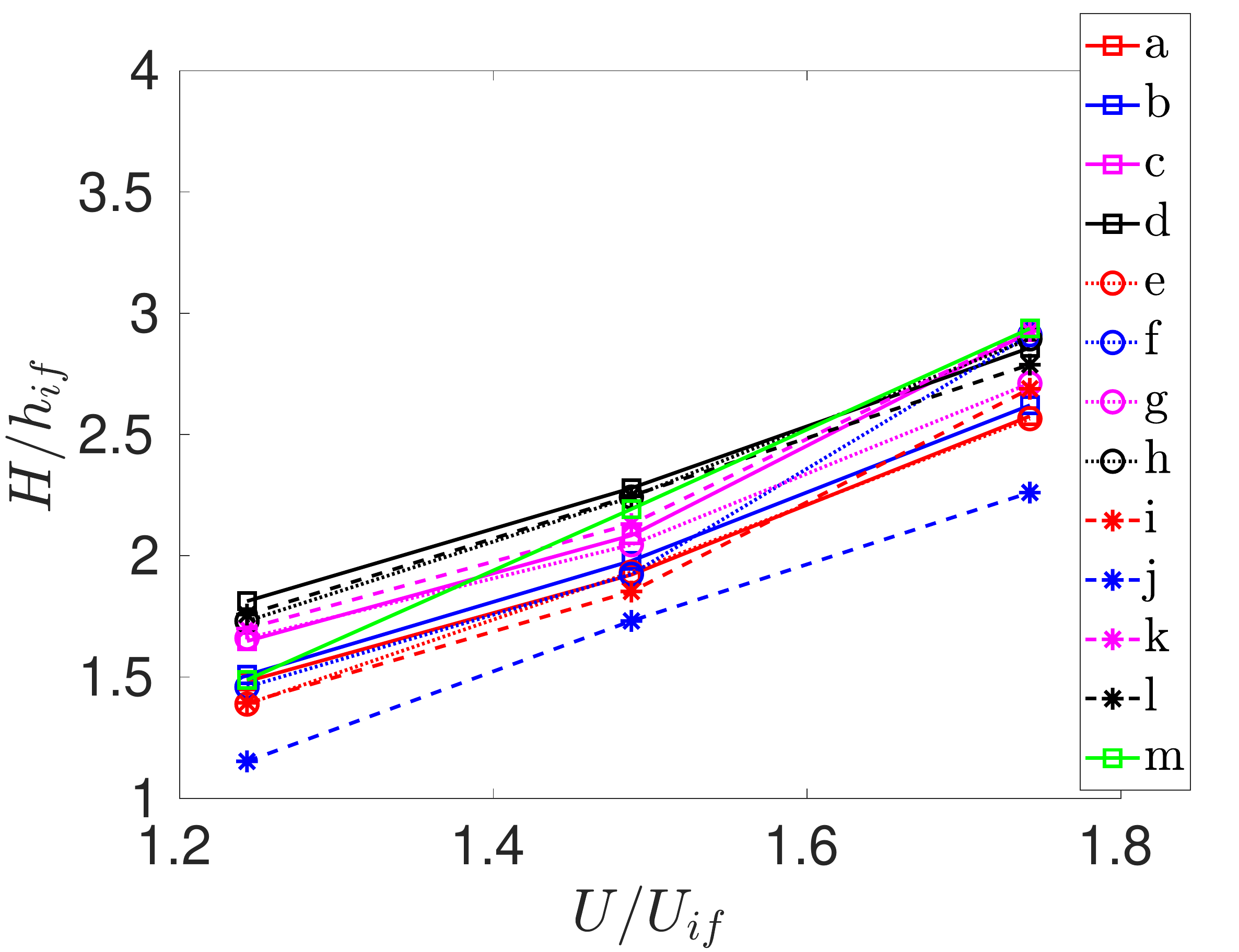}\\
			(b)
		\end{tabular}
	\end{minipage}
	\hfill
		\begin{minipage}{0.49\linewidth}
		\begin{tabular}{c}
			\includegraphics[width=0.85\linewidth]{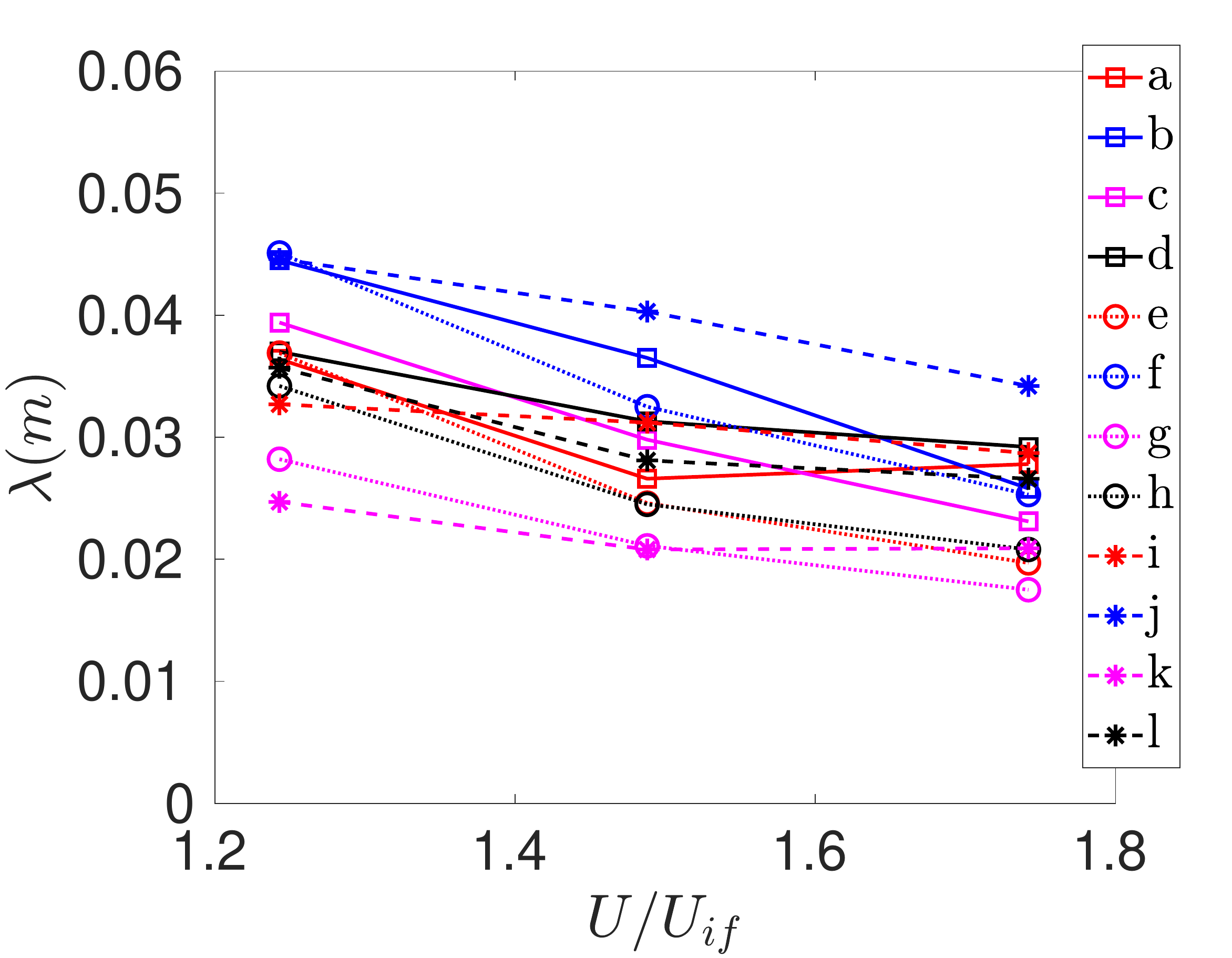}\\
			(c)
		\end{tabular}
	\end{minipage}
	\hfill
	\begin{minipage}{0.49\linewidth}
		\begin{tabular}{c}
			\includegraphics[width=0.85\linewidth]{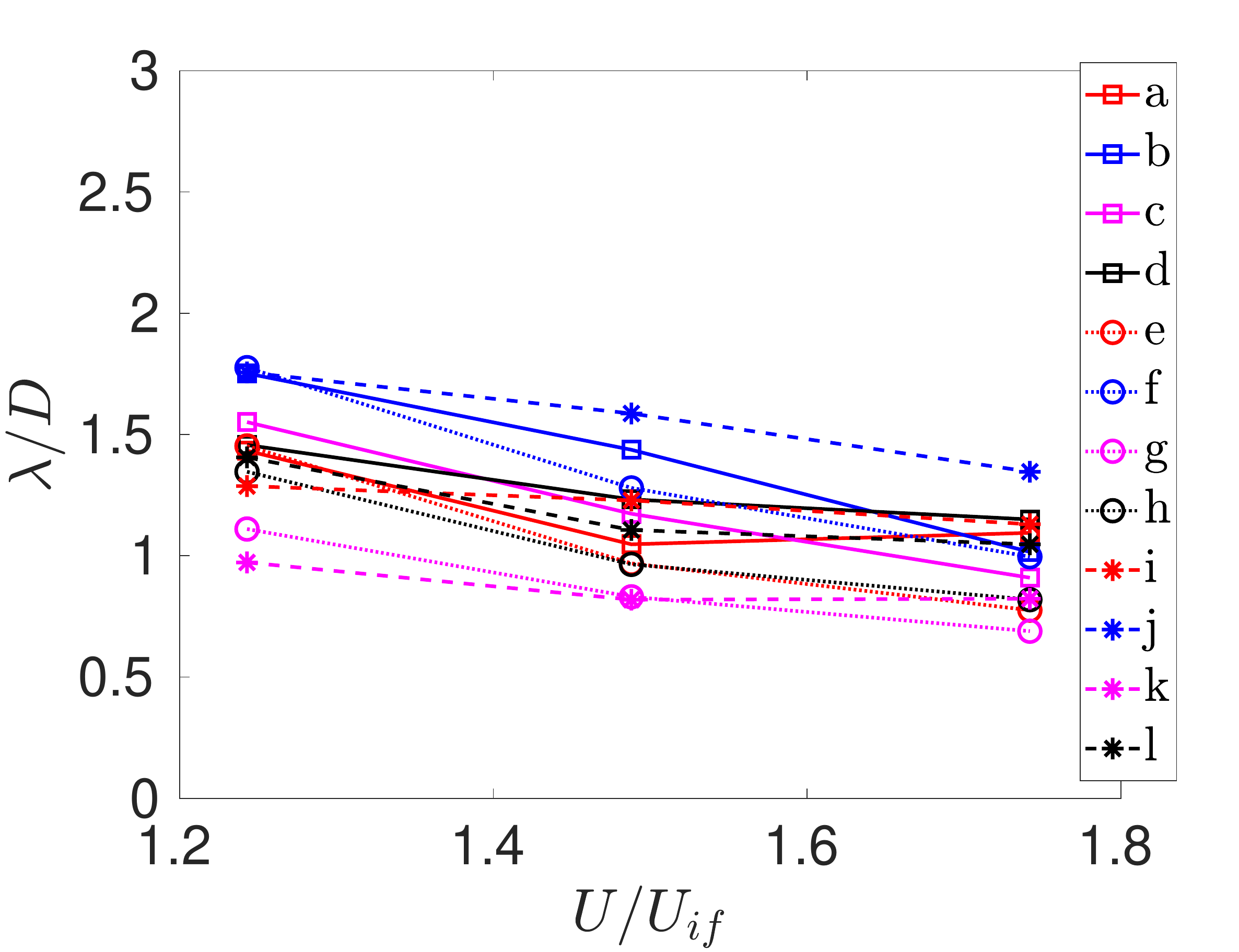}\\
			(d)
		\end{tabular}
	\end{minipage}
	\hfill
	\caption{Bed height $H$ and plug length $\lambda$ from numerical simulations as functions of the fluid velocity $U$ normalized by $U_{if}$, for different density of grains (grain material) and number of elements. Figures (a) and (b) show the dimensional and dimensionless bed heights, $H$ and $H/h_{if}$, respectively, and figures (c) and (d) dimensional and dimensionless plug lengths, $\lambda$ and $\lambda /D$, respectively. Each symbol corresponds to a different case listed in Tab. \ref{table:table4}}
	\label{fig:variation_density}
\end{figure}

\begin{table}[h!]
\caption{Numerical simulations varying the number of elements $N$ and density of grains $\rho_p$. The table presents the label of cases as in Fig. \ref{fig:variation_density}, type of element, number of elements $N$, and density of grains $\rho_p$ normalized by that of water $\rho_f$.}
\label{table:table4}
\centering
\begin{tabular}{c c c c}  
\hline\hline
Case & Type & $N$ & $\rho_p/\rho_f$\\
\hline 
a & Trio & 100 & 2.76\\
b & Trio & 200 & 2.76\\
c & Duo & 150 & 2.76\\
d & Duo & 300 & 2.76\\
e & Trio & 100 & 2.50\\
f & Trio & 200 & 2.50\\
g & Duo & 150 & 2.50\\
h & Duo & 300 & 2.50\\
i & Trio & 100 & 1.19\\
j & Trio & 200 & 1.19\\
k & Duo & 150 & 1.19\\
l & Duo & 300 & 1.19\\
m & Trio & 300 & 2.76\\
\hline
\hline 
\end{tabular}
\end{table}

Considering the good agreement between the results from numerical simulations and those from experiments, we carried out simulations varying the density of particles $\rho_p$ and the number of elements $N$ in order to investigate their effects on the average bed height and plug length. Besides the density of aluminum, $\rho_p/\rho_f$ = 2.76, we used those of sand and organic matter, $\rho_p/\rho_f$ = 2.50 and 1.19, respectively. We observed the same behavior for all densities and number of grains, with the formation of granular plugs and oscillation of the top of the bed, and a summary of the numerical results can be seen in Fig. \ref{fig:variation_density}. Figures \ref{fig:variation_density}a and \ref{fig:variation_density}b show the dimensional and dimensionless bed heights, $H$ and $H/h_{if}$, respectively, and Figs. \ref{fig:variation_density}c and \ref{fig:variation_density}d dimensional and dimensionless plug lengths, $\lambda$ and $\lambda /D$, respectively, as functions of the fluid velocity $U$ normalized by $U_{if}$, for different density of grains (grain material) and number of elements (the symbols are in the figure key and listed in Tab. \ref{table:table4}). We observe an increase in the bed height with the water velocity, heights going from 1.1$h_{if}$ $\lesssim$ $H$ $\lesssim$ 1.8$h_{if}$ to 2.2$h_{if}$ $\lesssim$ $H$ $\lesssim$ 2.9$h_{if}$ for $U$ going from 1.2 to 1.8$U_{if}$ for all particle densities and number of elements. For the same range of water velocities, the plug length decreases with $U$, going from 1$D$ $\lesssim$ $\lambda$ $\lesssim$ 2$D$ to 0.8$D$ $\lesssim$ $\lambda$ $\lesssim$ 1.5$D$ for all densities and number of elements. The graphics of Fig. \ref{fig:variation_density} show that the bed behavior remains roughly the same for different densities and number of grains, that the bed height scales with that at incipient fluidization while the plug length scales with the tube diameter, and that the fluid velocity scales with that at incipient fluidization.

\subsection{Motion of Grains}

\begin{figure}[ht]
	\begin{minipage}{0.49\linewidth}
		\begin{tabular}{c}
			\includegraphics[width=0.3\linewidth]{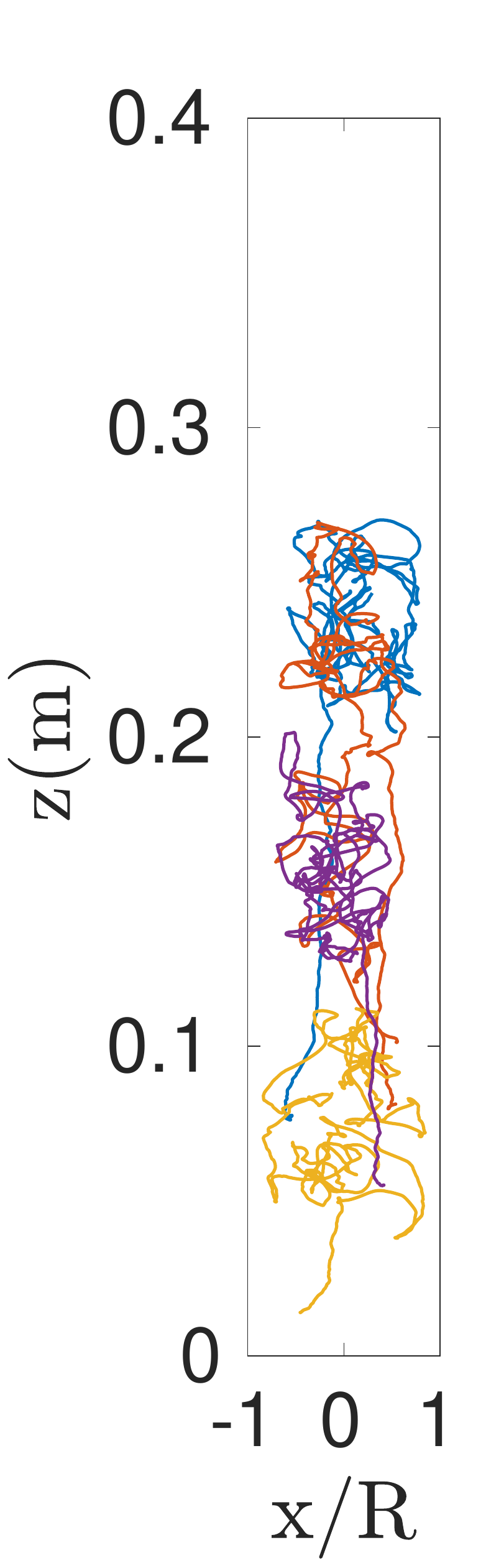}\\
			(a)
		\end{tabular}
	\end{minipage}
	\hfill
	\begin{minipage}{0.49\linewidth}
		\begin{tabular}{c}
			\includegraphics[width=0.3\linewidth]{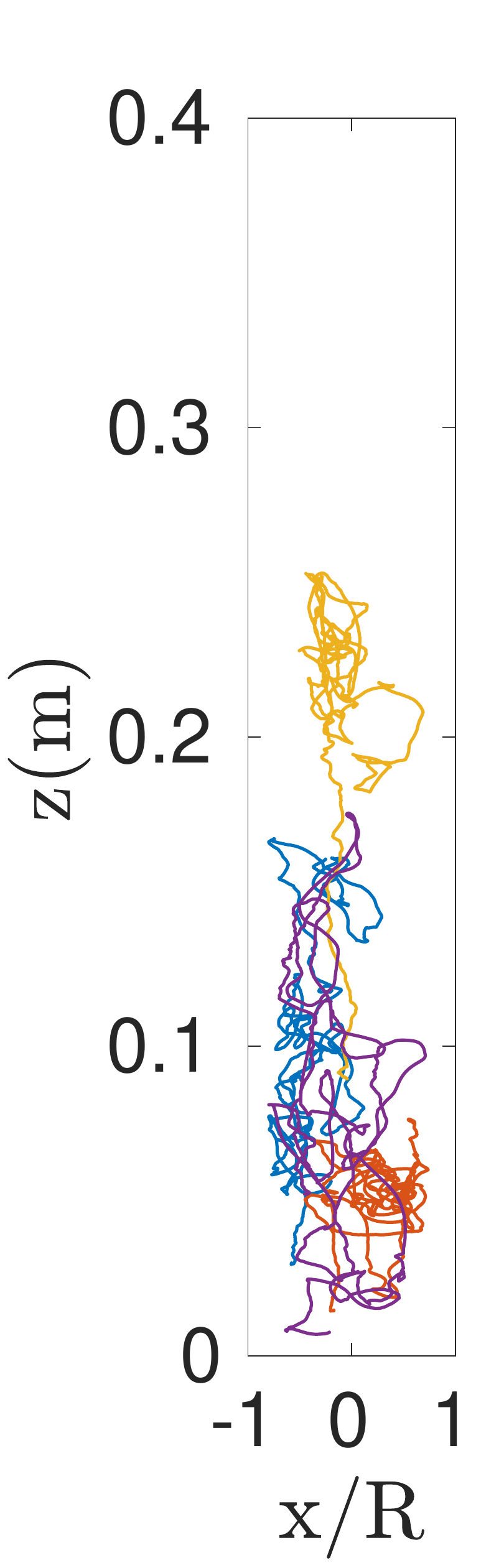}\\
			(b)
		\end{tabular}
	\end{minipage}
	\hfill
		\begin{minipage}{0.49\linewidth}
		\begin{tabular}{c}
			\includegraphics[width=0.3\linewidth]{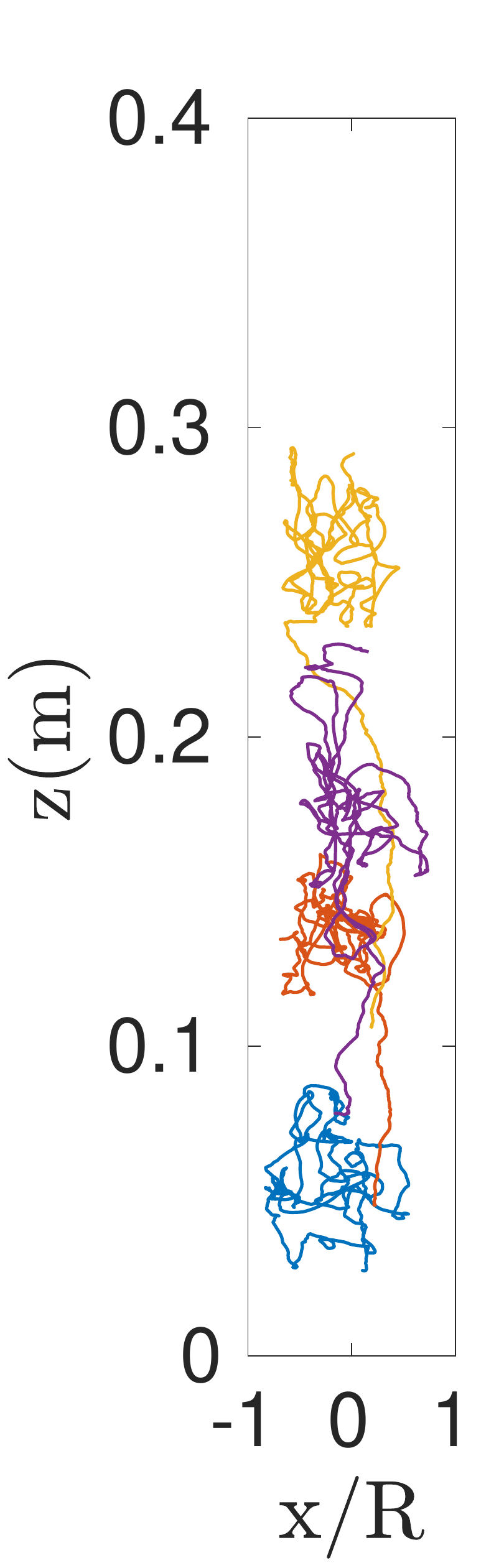}\\
			(c)
		\end{tabular}
	\end{minipage}
	\hfill
	\begin{minipage}{0.49\linewidth}
		\begin{tabular}{c}
			\includegraphics[width=0.3\linewidth]{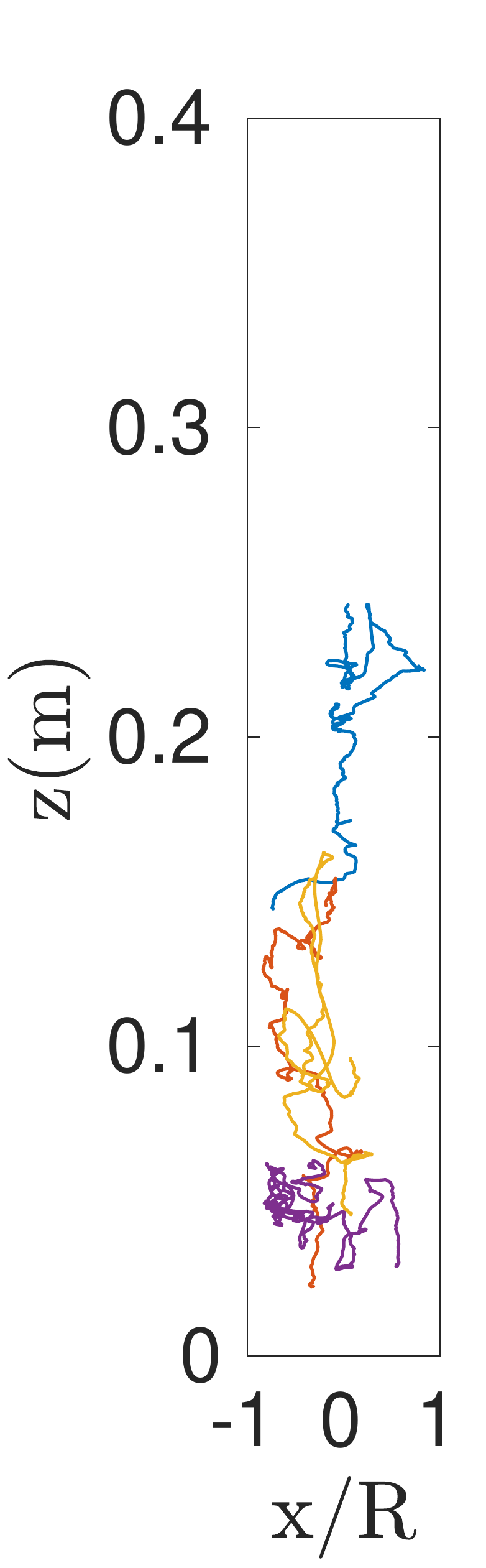}\\
			(d)
		\end{tabular}
	\end{minipage}
	\hfill
	\caption{Trajectories obtained numerically for the centroid of some (a) duos at $U$ = 0.164 m/s; (b) trios at $U$ = 0.164 m/s; (c) duos at $U$ = 0.137 m/s; and (d) trios at $U$ = 0.137 m/s. Total duration is 14 s, and each color (or gray scale) corresponds to a different element.}
	\label{fig:trajectories}
\end{figure}

The trajectories obtained from experiments are limited to particles in contact with the tube wall and small durations, since the bed is opaque and computations of trajectories are interrupted when particles move deeper in the bed. For that reason, we present trajectories computed from experiments in the supplementary material, only as a reference, and use those from numerical simulations in our analysis. Some trajectories are illustrated in Fig. \ref{fig:trajectories}.

Figures \ref{fig:trajectories}a and \ref{fig:trajectories}c show four trajectories of individual duos for beds consisting of 300 elements and $U$ equal to 0.164 and 0.137 m/s, respectively, and Figs. \ref{fig:trajectories}b and \ref{fig:trajectories}d four trajectories of trios for beds consisting of 200 elements and $U$ equal to 0.164 and 0.137 m/s, respectively. From Fig. \ref{fig:trajectories}, we note that individual elements are bounded to specific regions in the vertical direction, different from what happens with loose spheres, where individual elements travel along the entire bed \cite{Cunez2}. This reduced mobility can be explained by the high confinement experienced by each bonded particle. In the case of trios, we observe a much lower mobility for $U$ = 0.137 m/s, with grains moving smaller distances when compared to duos. The stronger confinement of trios (the cross-sectional ratio between the tube and elements is reduced by a factor 9 with respect to loose spheres) hinders vertical displacements of each element.

Finally, we computed instantaneous rms averages of the norm of the velocity for all particles within beds, and one example is presented in the supplementary material for both experiments and numerical simulations. The magnitudes of averaged velocities and their time evolution show that the numerical outputs are in accordance with experimental results.

\subsection{Clustering and Jamming}
\label{subsection_jamming}

Instantaneous fluctuations around an ensemble average can be evaluated by using images from experiments since fluctuations are computed at each instant, being unnecessary to follow grains over long durations. We thus tracked along movie frames the grains in contact with the tube wall with numerical scripts based on Kelley and Ouellette \cite{Kelley} and Houssais et al. \cite{Houssais_1}, and computed their fluctuations. The velocity fluctuations in $x$ and $y$ directions (Fig. \ref{fig:1}), $u_p$ and $v_p$, respectively, were computed as deviations of the instantaneous velocities of each grain from the average value for the ensemble. From the experimental data, the two-dimensional granular temperature $\theta$ was computed as in Eq. \ref{Eq:grain_temp_exp}:

\begin{equation}
\theta \,=\, \frac{1}{2} \left( u_p^2+v_p^2 \right).
\label{Eq:grain_temp_exp}
\end{equation}

\begin{figure}[ht]
	\begin{center}
	\begin{tabular}{c}
	\includegraphics[width=0.8\columnwidth]{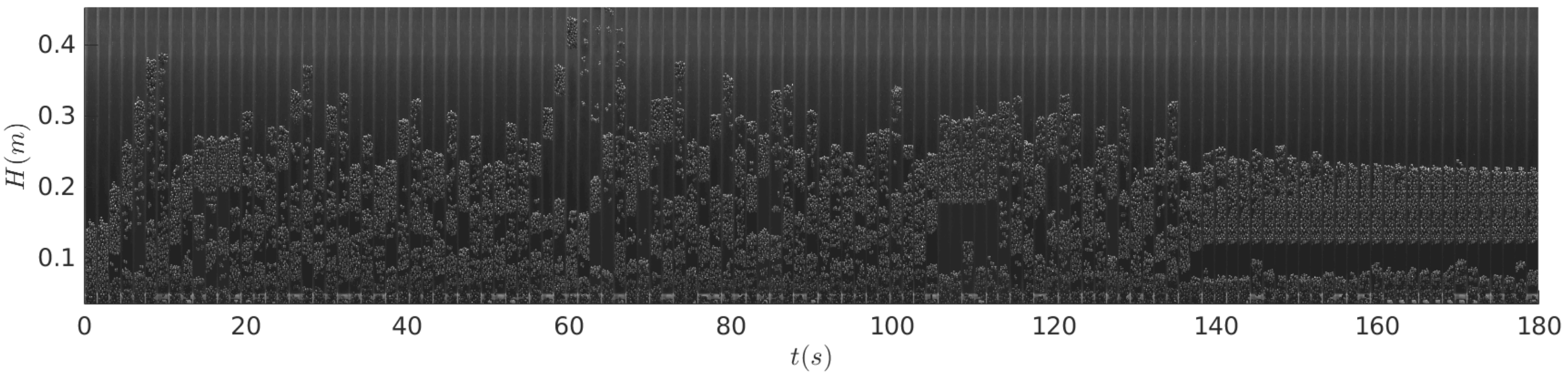}\\
	(a)\\
	\includegraphics[width=0.65\columnwidth]{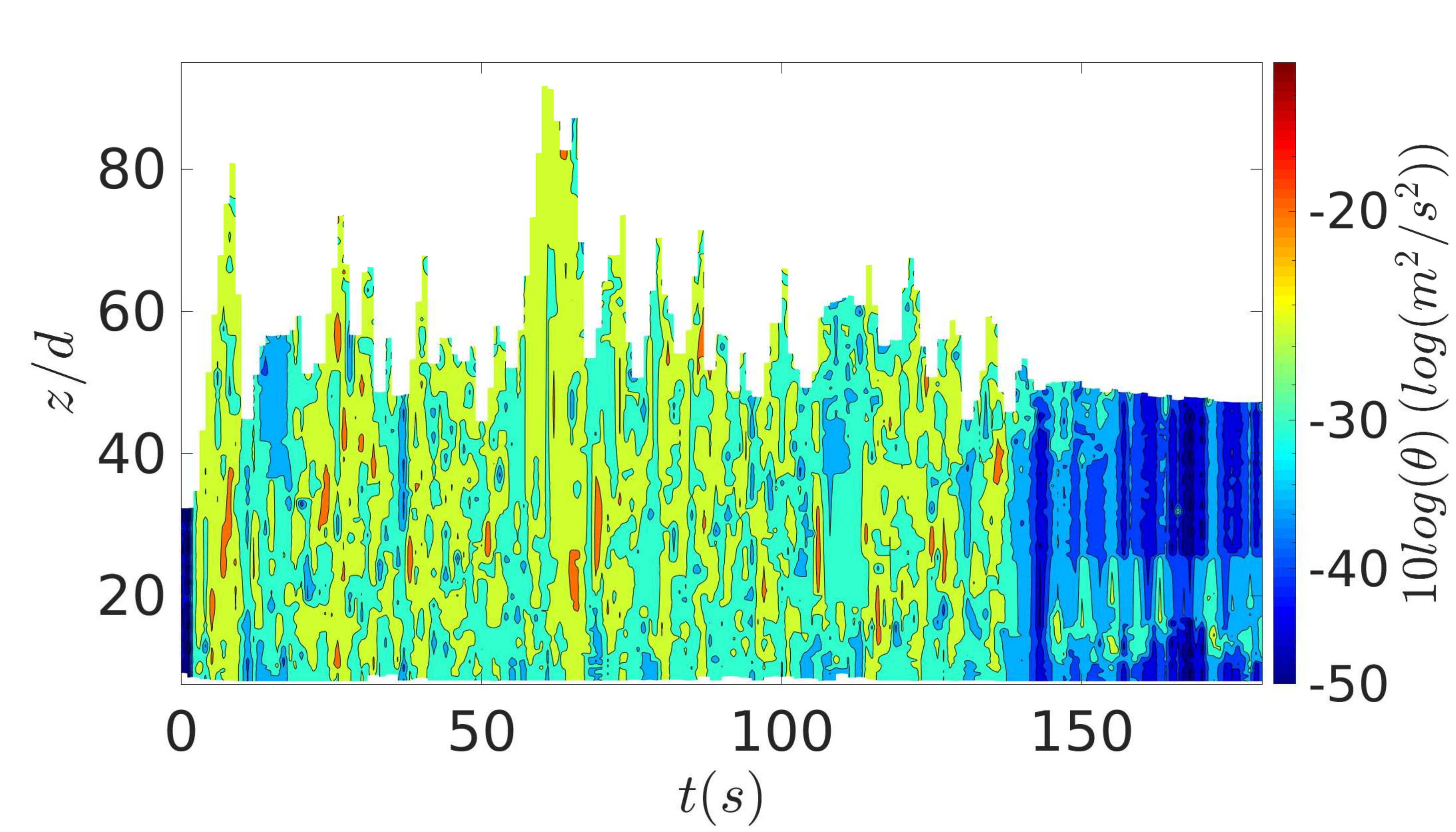}\\
	(b)
	\end{tabular} 
\end{center}
	\caption{(a) Snapshots placed side by side of particle positions for beds consisting of trios and $U$ = 0.137 m/s, with total time of 180 s and time between frames of 1.5 s. (b) Spatio-temporal diagram of cross-sectional averages of the granular temperature computed from movie frames.}
	\label{fig:snapshots_jamming}
\end{figure}

Figure \ref{fig:snapshots_jamming}a shows snapshots of particle positions, until the instant $t$ = 180 s, for the bed consisting of trios presented in Fig. \ref{fig:snapshots_trio}c, and from which a strong blockage can be observed by the end of the sequence. In order to investigate if grains within the clogging structure are jammed, we estimated the instantaneous values of cross-sectional averages of the granular temperature by computing horizontal averages of $\theta$ for each movie frame. The averaged values of $\theta$ are presented in Fig. \ref{fig:snapshots_jamming}b, where 10$\log \theta$ was plotted instead of $\theta$ in order to accentuate differences. From Fig. \ref{fig:snapshots_jamming}a, we observe that a clogging structure appears around $t$ $\approx$ 140 s, at the beginning with some fluctuations at the grain scale, and that from $t$ $\approx$ 160 s grains seem motionless and jammed. This is confirmed in Fig. \ref{fig:snapshots_jamming}b, where we observe lower values of $\theta$ from $t$ $\approx$ 160 s on, indicating small levels of microscopic motion and a jammed state \cite{Cunez3}.

\begin{figure}[ht]
	\begin{center}
	\begin{tabular}{c}
	\includegraphics[width=0.65\columnwidth]{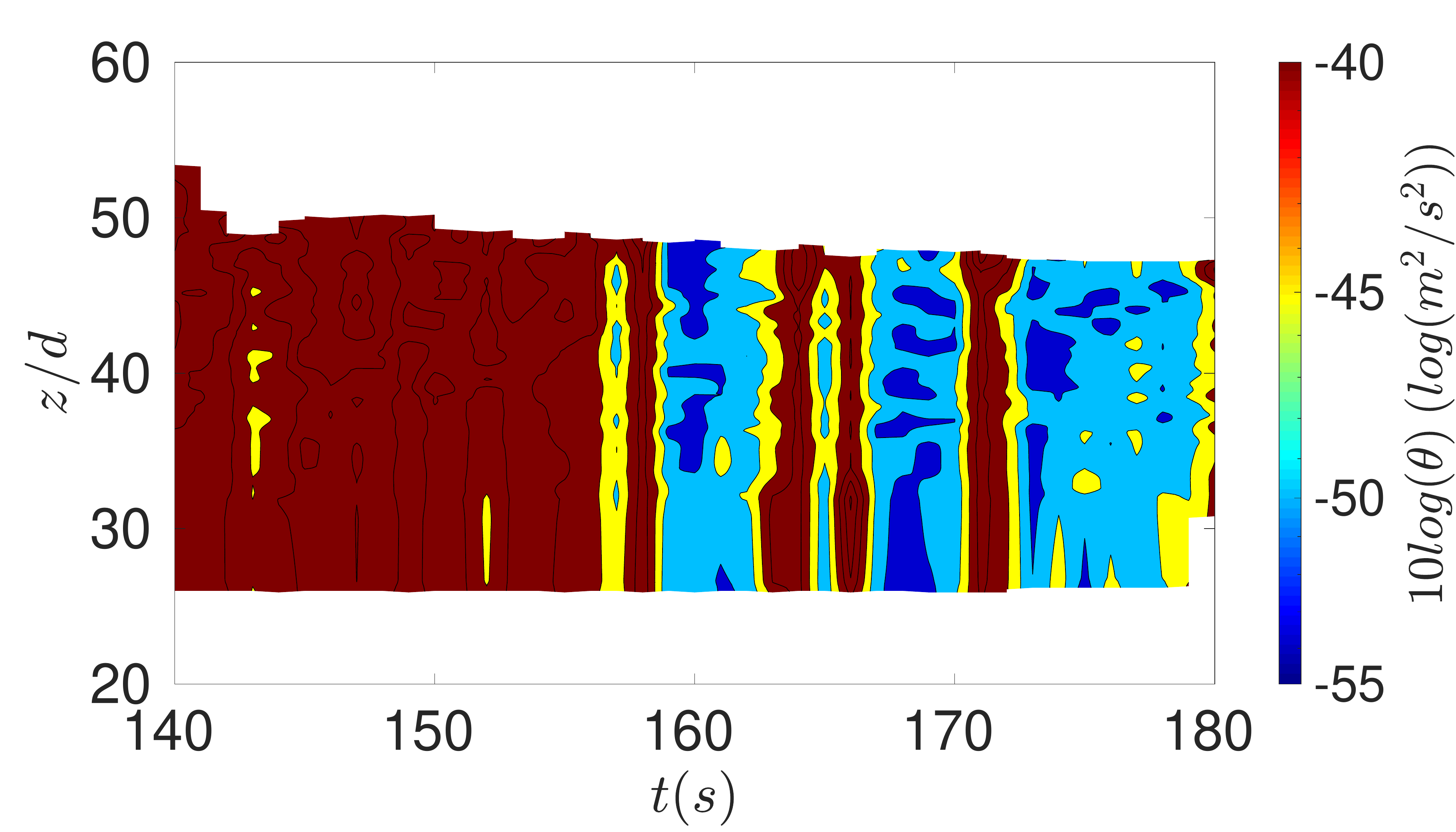}\\
	(a)\\
	\includegraphics[width=0.35\columnwidth]{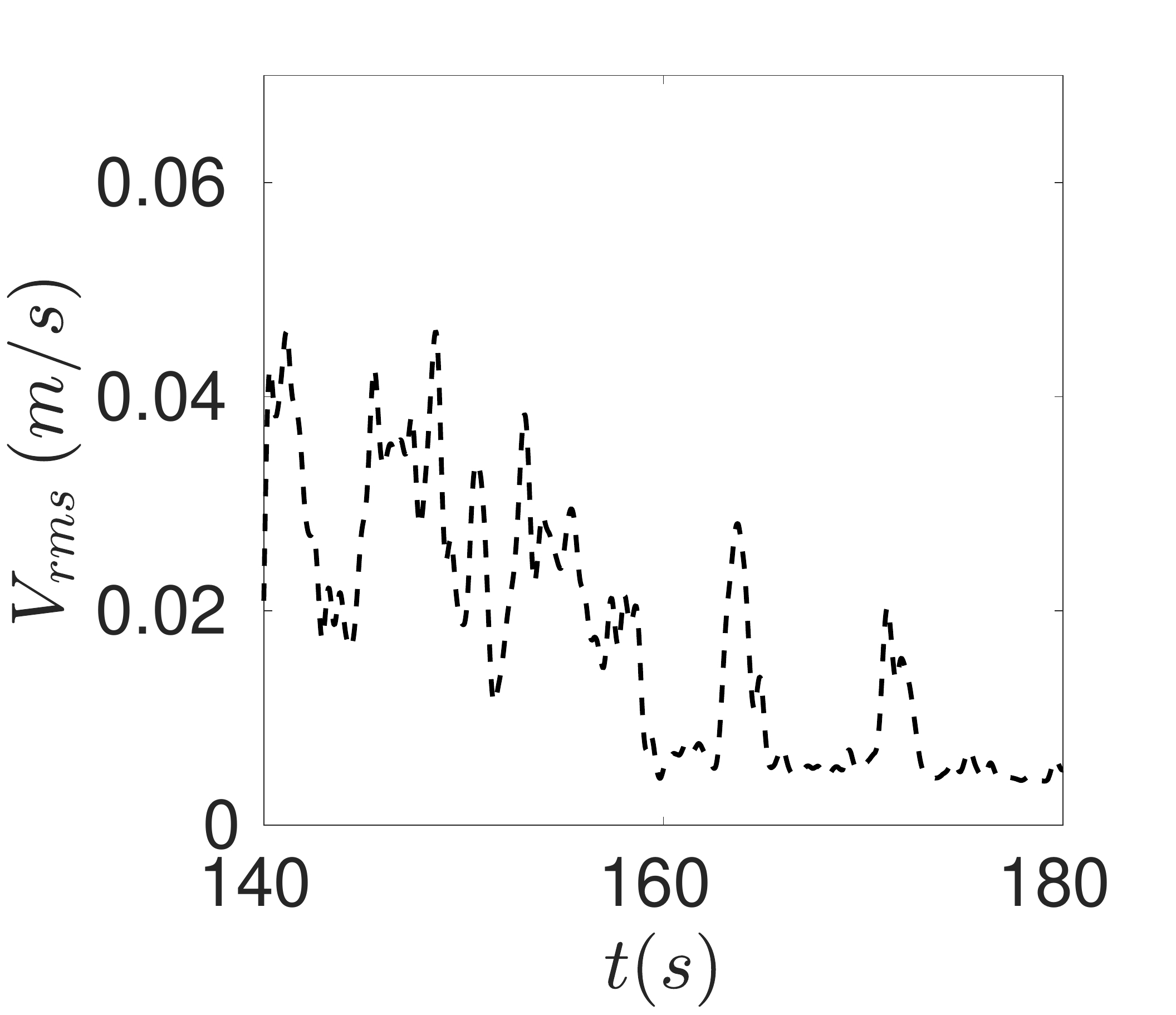}\\
	(b)
	\end{tabular}
\end{center}
	\caption{(a) Spatio-temporal diagram of cross-sectional averages of the granular temperature for the clogging structure. (b) Evolution of the rms average of the norm of the velocity for all particles in the clogging region.}
	\label{fig:jamming}
\end{figure}

We focus next on the clogging structure for a deeper analysis of its behavior. For that, we computed a rms average of the norm of the velocity (in the $xy$ plane), as in Ref. \cite{Cunez3}, given by Eq. \ref{Eq:rms_vel},

\begin{equation}
V_{rms} \,=\, \sqrt{\sum_{i=1}^N \frac{1}{N} \left( U_p \right) ^2 + \sum_{i=1}^N \frac{1}{N} \left( V_p \right) ^2},
\label{Eq:rms_vel}
\end{equation}

\noindent where $V_{rms}$ is the instantaneous rms average for the ensemble of grains within the clogging structure, $i$ refers to the $i^{th}$ grain, $N$ is the number of considered grains, and  $U_p$ and $V_p$ are the $x$ and $y$ components of the instantaneous velocity of each grain, respectively. Figures \ref{fig:jamming}a and \ref{fig:jamming}b present, respectively, the spatio-temporal diagram of cross-sectional averages of $\theta$ and the evolution of $V_{rms}$, computed for the clogging structure seen in Fig. \ref{fig:snapshots_jamming}. We observe from Figs. \ref{fig:jamming}a and \ref{fig:jamming}b the existence of intervals where the level of fluctuation is very low, corresponding to jammed states \cite{Cunez3}. These intervals alternate with others of higher fluctuation levels, showing that jammed states are intermittent, similar as happens for loose spheres \cite{Cunez3}; however, for bonded spheres jamming occurs at higher velocities and vertical positions in the tube (at approximately 10-20$D$ from the tube entrance). The jammed state persists unless the fluid velocity varies above or below given levels (not investigated in detail here). The explanation for jamming is that, once particles are packed together in a given region, forces are redirected from the vertical direction (drag and pressure forces) toward the horizontal direction by means of contacts, so that grains remain blocked and jamming occurs. This effect is stronger for more confined beds (case of trios).

In summary, a granular plug is frequently blocked in the middle of the tube and, after some time elapses, the trios are jammed with a very small level of fluctuation. In practical applications, such as fluidized beds in biological reactors, a jammed state of part of the bed can prevent the reactor operation. The present results shed some light on where and in which cases jamming appears. In particular, practical applications employing narrow beds may allow particle agglomeration in duos, but shall avoid the formation of trios since these elements present smaller displacements and the bed mobility is significantly reduced, with blockages occurring at 10-20$D$ from the tube inlet. In case of operational problems, this region should be inspected for the presence of clogs. From a more fundamental point of view, our results call into question the fluidization conditions, since even the microscopic motion of beds stops when jamming occurs.

\section{CONCLUSIONS}

In this paper we investigated the dynamics of solid-liquid fluidized beds of bonded spheres in very narrow tubes. For that, we carried out experiments and CFD-DEM simulations for beds consisting of duos and trios of bonded aluminum spheres with diameter of 4.8 mm fluidized by water in a 25.4 mm-ID pipe. We varied the water velocities and size of beds, which consisted of either 150-300 duos or 100-200 trios, and obtained experimental images that were processed at both the macroscopic and microscopic scales and numerical outputs with the instantaneous positions of all grains. Our results showed different structures within the bed and distinct motions of individual duos and trios. We observed the formation of plugs that appear for both duos and trios with length $\lambda$ of the order of the tube diameter, and that $\lambda$ decreases with increasing the water velocity, similar to what has been observed for loose spheres; however, the bonded elements are bounded to specific regions in the vertical direction, different from what happens in beds of loose spheres, where individual elements travel along all the bed \cite{Cunez2}. We conjecture that the lower mobility of bonded spheres is due to their stronger confinement, the cross-sectional ratio between tube and grains being reduced by a factor 9 in the case of trios, which hinders vertical displacements of individual elements. In the specific case of trios, the results showed that a granular plug is frequently blocked in the middle of the tube and, by analyzing the microscopic motion within the plug, we found that its elements are jammed after some time has elapsed, with virtually no motion. In applications using biological fluidized beds, a jammed state of part of the bed prevents the correct operation of the reactor. Our results shed some light on where and in which cases jamming appears, and, from a more fundamental point of view, call into question the fluidization conditions in the case of trios.

\section*{SUPPLEMENTARY MATERIAL}
See the supplementary material for microscopy images of the used grains, instantaneous snapshots of particle positions, and a table containing dimensional data from experiments and simulations and their respective standard deviations.

\section*{DATA AVAILABILITY}
The data that support the findings of this study are openly available in Mendeley Data at https://data.mendeley.com/datasets/pg4xyzz86n, Ref. \cite{Supplemental2}.

\begin{acknowledgments}
The authors are grateful to FAPESP (Grant Nos. 2016/18189-0, 2018/23838-3, 2018/14981-7 and 2019/20888-2) and to CNPq (grant no. 400284/2016-2) for the financial support provided.
\end{acknowledgments}

\bibliography{references}

\end{document}